\documentclass[12pt]{article}
\usepackage{amsfonts}
\usepackage{amsmath,amssymb}
\usepackage{mathrsfs}
\usepackage{array}
\usepackage{color}
\usepackage{amsxtra}
\usepackage{amstext}
\usepackage{latexsym}
\newtheorem{thm}{Theorem}[section]
\newtheorem{prop}[thm]{Proposition}
\newtheorem{lem}[thm]{Lemma}
\newtheorem{remark}[thm]{Remark}

\newtheorem{definition}[thm]{Definition}
\newtheorem{corollary}[thm]{Corollary}
\numberwithin{equation}{section}
\def\Tr{{\rm Tr\,}}

\def\Det{{\rm Det}}
\def\R{\Bbb R}
\def\C{\Bbb C}

\def\supp{{\rm supp\,}}

\usepackage{color}

\newcommand{\rd}{\textcolor{red}}

\begin{document}

\phantom{.} {\qquad \hfill \textit{\textbf{\rd{Draft: September 05, 2009 }}}}

\vskip 2.5cm

\setcounter{footnote}{0}
\renewcommand{\thefootnote}{\arabic{footnote}}

\begin{center}

{\Large{\bf {Large Deviation Principle for Non-Interacting Boson Random Point
{Processes}}}}

\vskip 1cm

{\textbf{Hiroshi Tamura} \footnote{tamurah@kenroku.kanazawa-u.ac.jp}\\
       Graduate School of the Natural Science and Technology\\
           Kanazawa University,\\
          Kanazawa 920-1192, Japan

\bigskip

\textbf{Valentin A.Zagrebnov }\footnote{zagrebnov@cpt.univ-mrs.fr}\\
Universit\'e de la M\'editerran\'ee(Aix-Marseille II) and  \\
Centre de Physique Th\'eorique - UMR 6207 \\
Luminy-Case 907, 13288 Marseille Cedex 9, France}

\vspace{1.5cm}

\end{center}
%%%%%%%%%%%%%%%%%%%%%%%%%%%%%%%%%%%%%%%%%%%%%%%%%%%%%%%%%%%%%%%%%%%%%%%%%%%%

\begin{abstract}
\noindent Limit theorems, including the large deviation principle,
are established for random point processes (\textit{fields}), which describe the position
distributions of the perfect boson gas in the regime of the Bose-Einstein
condensation. We compare these results with those for the case of the \textit{normal} phase.
\end{abstract}

\vspace{1.5cm}


\newpage
%%%%%%%%%%%%%%%%%%%%%%%%%%%%%%%%%%%%%%%%%%%%%%%%%%%%%%%%%%%%%%%%%%%%%%%%%%%%%%%%%%%%%%
\tableofcontents

%\newpage
%%%%%%%%%%%%%%%%%%%%%%%%%%%%%%%%%%%%%%%%%%%%%%%%%%%%%%%%%%%%%%%%%%%%%%%%%%

%%%%%%%%%%%%%%%%%%%%%%%%%%%%%%% INTRODUCTION %%%%%%%%%%%%%%%%%%%%%%%%%%%%%%%%%%%%%%%%%%
\section{Introduction and Main Results}
%%%%%%%%%%%%%%%%%%%%%%%%%%%%%%%%%%%%%%%%%%%%%%%%%%%%%%%%%%%%%%%%%%%%%%
Fermion and boson random point processes (fields, or general \textit{Cox processus}) were studied
by many authors,
in particular since they have a deep connection with the quantum statistical mechanics
\cite{1,3,4, F, Fr, FFr1, FFr2}. See also \cite{S, Ly} and references therein.
One of the advantages of the random point field approach to quantum statistical mechanical models
is that it enables probabilistic limit theorems to apply to these models.
In \cite{ST03}, typical limit theorems are given for a certain class of random point processes
which include the particular cases of the fermion as well as boson random point processes.
In \cite{2}, the random point processes, which describe the position distribution of constituent
particles of boson gases in Bose-Einstein Condensation (BEC) are constructed for the first time.

The purpose of the present paper is to give the limit theorems, such as the Law of
the Large Numbers (LLN), the Central Limit Theorem (CLT) and the Large Deviation Principle (LDP)
for the
random boson point processes in the regime of the BEC. We compare them
with the corresponding theorems for the normal phase (i.e. without the BEC). In the latter case
a detailed study of the limit theorems, which do not use the random point processes formalism,
is due to \cite{LLS} and \cite{GLM}. In the last reference the authors consider even
interacting quantum gases, but only in the rarified regime insuring the normal phase. These papers
motivated the study of the large deviation principle in the Bogoliubov-type models \cite{BZ},
where BEC plays a key r\^{o}le in description of the model thermodynamic behaviour and the spectrum
of excitations.

The study of the boson random point processes in the BEC regime is an interesting and
delicate mathematical problem \cite{2}, see also a recent paper \cite{E}.
This last paper makes evident that a Cox process in the BEC regime \cite{2} is driven by the square
norm of a shifted Gaussian process. The shift is particle density dependent. In particular, this
observation makes a contact with the Dynkin isomorphism theorem (known for Gaussian processus)
as well as
a relation between infinite devisability  and factorisation of the boson Cox process involved
in the BEC.

In the present paper, we study the limit theorems in the BEC regime
( Theorems \ref{LLN}, \ref{CLTh}, \ref{LDP}) and discuss in Conclusion the comparison with
the analogue of these Theorems in the normal phase.

%%%%%%%%%%%%%%%%%%%%%%%%%%%%%%%%% Results %%%%%%%%%%%%%%%%%%%%%%%%%%%%%%%%%%%%%%%%%%%%%%%%%%
Let $ \{G:= \exp(\beta\Delta)\}_{\beta\geq 0} $ be the (set-adjoint) \textit{heat semigroup}
generated by the Laplacian acting in $L^2(\R^d)$.
For any non-negative bounded measurable function $f\geq 0$ with a compact support in $\R^d$ ,
the operator
\[
         W_f := (G(1-G)^{-1})^{1/2}\sqrt{1-e^{-f}} \ ,
\]
is a bounded and
\[
        K_f := W_f^*W_f \in {\mathfrak{C}_{1}(L^2(\R^d))} \ ,
\]
i.e., is a \textit{trace-class} operator on $L^2(\R^d)$. If $f=0$, then the operator $K:=K_{f=0}$ is
bounded with the \textit{translation-invariant} kernel:
\begin{equation*}
K(x,y) = \int_{\R^d}\frac{dp}{(2\pi)^d} \, \frac{e^{i (x-y)p}}{e^{\beta|p|^2}-1}  \ .
\end{equation*}

Below we consider a noninteracting boson random point field $\nu_{\rho}$ for the total particle density
$ \rho > \rho_c$, which is characterized by the generating functional \cite{2}:
\begin{equation}\label{gen-func-BEC}
       \int_{Q(\R^d)} e^{-\langle f, \xi \rangle} \nu_{\rho}(d\xi)
    = \frac{ \exp(-(\rho-\rho_c)( \sqrt{1-e^{-f}}, (1+K_f)^{-1}\sqrt{1-e^{-f}}))}
           { \Det[1+K_f]  } \ .
\end{equation}
Here $Q(\R^d)$ is the space of all point measures on $\R^d$, $\Det$ stands for the \textit{Fredholm
determinant} and $\langle f, \xi \rangle = \sum_{j} f(x_j)$, if $\xi= \sum_{j} \delta_{x_j} \in Q(\R^d)$.

The \textit{critical} particle density, $\rho_c := \rho_{c}(\beta)$, for the
\textit{perfect} Bose-gas can be expressed as:
\[
         \rho_{c}(\beta) := K(x,x) =
          \int_{\R^d} \frac{dp}{(2\pi)^d} \, \frac{e^{-\beta|p|^2}}{1-e^{-\beta|p|^2}}
             = \frac{\zeta(d/2)}{(4\pi\beta)^{d/2}} \ .
\]

The random point field $\nu_{\rho}$ defined by by the generating functional (\ref{gen-func-BEC})
was introduced in \cite{2} to describe the Bose-Einstein condensation in the non-interacting (perfect)
boson gas. For the detailed presentation of these notions, we refer to Ref.\cite{2}.
(See also the next Section \ref{2}.)

Below in the present paper, we use the following notations: $\| \, \cdot \, \|_p$ for $L^p(\R^d)$
norm and $\| \, \cdot \, \|$ for the bounded operators norm on $L^2(\R^d)$.

With these notations the main results of the paper can be expressed as follows:
%%%%%%%%%%%%%%%%%%%%%%%%%%%%%%%%%%%%%%%%%%%%%%%%%%%%%%%%%%%%%%%%%%%%%%%%%%%%%%%%%%%%%%%%%%%%%%
\begin{thm}[Law of Large Numbers]\label{LLN}  For $\kappa \to \infty$, the limit
\[
      \frac{1}{\kappa^d}\langle  f(\cdot/\kappa), \xi \rangle \longrightarrow
      \rho \int_{\R^d}  dx\, f(x)
\]
holds in $L^2( Q(\R^d), \nu_{\rho})$ .
\end{thm}
%%%%%%%%%%%%%%%%%%%%%%%%%%%%%%%%%%%%%%%%%%%%%%%%%%%%%%%%%%%%%%%%%%%%%%%%%%%%%%%%%%%%%%%%%%%%%%%
\begin{thm}[Central Limit Theorem]\label{CLTh} Let
\[
      Z_{\kappa} := \frac{\langle  f(\cdot/\kappa), \xi \rangle -
                  \kappa^d \rho \int_{\R^d} dx\, f(x)}
          {\sqrt{2(\rho-\rho_c)} \, \|(-\beta\Delta)^{-1/2}f \|_2 \
\kappa^{(d+2)/2}} \ \ .
\]
Then the limit
\[
     \lim_{\kappa \to \infty}\int_{Q(\R^d)} e^{itZ_{\kappa}} \nu_{\rho}(d\xi)
           =  e^{-t^2/2} \ \ .
\]
\end{thm}
%%%%%%%%%%%%%%%%%%%%%%%%%%%%%%%%%%%%%%%%%%%%%%%%%%%%%%%%%%%%%%%%%%%%%%%%%%%%%%%%%%%%%%%%%%%%%%%
\begin{thm}[Large Deviation Principle]\label{LDP}
There exists a certain (\textit{bona fide})  rate convex function $I:\R \mapsto [0, + \infty]$,
such that the limits
\[
     \limsup_{\kappa\to\infty}\frac{1}{\kappa^{d-2}} \log \nu_{\rho}
     \Big( \frac{1}{\kappa^d}\big\langle f\big(\cdot/\kappa\big),
    \xi\big\rangle \in F\Big) \leqslant -\inf_{s\in F}I(s)
     \qquad \hphantom{L}
\quad {\rm{for \ any \ closed \ }} F \subset \R \ ,
\]
and
\[
     \liminf_{\kappa\to\infty}\frac{1}{\kappa^{d-2}} \log \nu_{\rho}
     \Big( \frac{1}{\kappa^d}\big\langle f\big(\cdot/\kappa\big),
    \xi\big\rangle \in G\Big) \geqslant -\inf_{s\in G}I(s)
    \qquad \hphantom{L}
\quad {\rm{for \ any \ open \ }} G \subset \R \ ,
\]
hold.
\end{thm}
%%%%%%%%%%%%%%%%%%%%%%%%%%%%%%%%%%%%%%%%%%%%%%%%%%%%%%%%%%%%%%%%%%%%%%%%%%%%%%%%%%%%%%%%%%%%%
In  Section 5 we compare these results with those for boson random point processes in the
\textit{normal phase}.

%%%%%%%%%%%%%%%%%%%%%%%%%% Abstract theory %%%%%%%%%%%%%%%%%%%%%%%%%%%%%%%%%%%%%%%%%%%%%%%%%%%%%%%%%
\section{Preliminary arguments and general setting}\label{2}
%%%%%%%%%%%%%%%%%%%%%%%%%%%%%%%%%%%%%%%%%%%%%%%%%%%%%%%%%%%%%%%%%%%%%%%%%%%%%%%%%%%%%%%%%%%%%%%%%%%%
Let $R$ be a locally compact Hausdorff space with countable basis, and
$\lambda$ be a positive Radon measure on $R$.
We suppose that the non-negative (possibly unbounded) self-adjoint operator $K$
in $L^2(R, \lambda)$ satisfies the following condition {\bf K'}.

\smallskip

\noindent {\bf Condition K' }: \par
 {\sl  \hspace{-5mm}(i) (locally trace class) For every bounded Borel set $\Lambda \subset R$,
 $K^{1/2}\chi_{\Lambda}$ is a Hilbert-Schmidt
operator, where $\chi_{\Lambda}$ denotes the multiplication operator corresponding to
the indicator function of the set $\Lambda$, which we denote by the same symbol. \par
 \hspace{-5mm}(ii) The operator $G=K(1+K)^{-1}$ has non-negative integral kernel $G(x,y)$,
which satisfies the conditions:}
\begin{eqnarray*}
    & G(x,y) > 0 & \quad  \lambda\otimes\lambda -a.e. \, (x,y) \in R^2 \ ,
\\
   & \int_RG(x,y)\, \lambda(dy) \leqslant 1   & \quad  \lambda -a.e. \, x \in R \ .
\end{eqnarray*}

\medskip

The above conditions are arranged in such a way that one can simultaneously deal with the
random point processes $\mu_K^{det}$ and $\mu_{K, \rho}$ , see \cite{2} and \cite{ST03}.
In particular, the operator $K$ has a positive kernel $K(x,y)$, i.e.,
\[
         K(x, y) > 0 \qquad \lambda\otimes\lambda -a.e. \, (x,y) \in R^2 \ ,
\]
see \cite{2}. The operator $K_{\Lambda}:=(K^{1/2}\chi_{\Lambda})^*K^{1/2}\chi_{\Lambda}$
is a \textit{trace-class} operator.
For a bounded measurable function $f$ with compact support, we define the operator
\[
        K_f := \sqrt{1-e^{-f}}K_{\Lambda} \sqrt{1-e^{-f}},
\]
where supp$\, f \subset \Lambda$.
Note that $K_f$ is independent of the choice of $\Lambda$, which contains supp$\, f$.

\medskip

Let $Q(R)$ be Polish space of all locally finite non-negative integer-valued Borel measures on $R$.
Recall that the Borel probability measures on $Q(R)$ (i.e. random point processes on $R$)
$\mu^{(det)}_K$ and $\mu_{K, \rho}$  are introduced in \cite{ST03, 2} for $\rho > 0$ by means of
generating
functionals:
\begin{equation}
   \int_{Q(R)}e^{-\langle f, \xi \rangle} \mu_K^{(det)}(d\xi)
   = \Det[1+K_f]^{-1} = \Det[1+(1-e^{-f})K_{\Lambda}]^{-1} \ ,
\label{chdet}
\end{equation}
\begin{eqnarray}
\int_{Q(R)}e^{-\langle f, \xi \rangle} \mu_{K,\rho}(d\xi)
&=& \exp\{-\rho\langle \sqrt{1-e^{-f}}, \, (1+K_f)^{-1}\sqrt{1-e^{-f}} \, \rangle \}  \nonumber \\
&=& \exp\{-\rho\langle \chi_{\Lambda}, \, (1+(1-e^{-f})K_{\Lambda})^{-1}(1-e^{-f})\rangle\} \ ,
\label{chrho}
\end{eqnarray}
for any function $f$ with $\Lambda \supset $ supp$ \, f$.

It was shown \cite{2} that for $ R = \R^d$ the boson random point processes corresponding  to the
ideal Bose-gas in the regime of Bose-Einstein condensation ($\rho > \rho_c$) is described by the
\textit{convolution}
$ \nu_{\rho} := \mu_K^{det}*\mu_{K, \, \rho-\rho_c}$.

%%%%%%%%%%%%%%%%%%%%%%%%%%%%%%%%%%%%    thmexp   %%%%%%%%%%%%%%%%%%%%%%%%%%%%%%%%%%%%%%%%%%%%%%%%%%%
\begin{thm}
For any non-negative bounded measurable function $f$ on $R$ with compact support
supp$\, f \subset \Lambda$
in a bounded Borel set  $\Lambda \subset R$ one has
the following equalities:
\begin{eqnarray*}
{\rm{(1)}} \ \hskip0.5cm
    \int_{Q(R)}e^{i\langle f, \xi \rangle} \mu_{K,\rho}(d\xi)
      &=& \exp[-\rho\big\langle \chi_{\Lambda}, \, (1+(1-e^{if})K_{\Lambda})^{-1}(1-e^{if})
    \rangle], \\
{\rm{(2)}}  \ \ \hskip0.5cm
    \int_{Q(R)}e^{\langle f, \xi \rangle} \mu_{K,\rho}(d\xi)
      &=& \begin{cases}
         \exp[\rho\langle \sqrt{e^{f}-1}, \, (1- \sqrt{e^{f}-1}K_{\Lambda}\sqrt{e^{f}-1})^{-1}
           \sqrt{e^{f}-1} \, \rangle ] < \infty & \hphantom{KKKKK} \\
  \hphantom{KKKKKKKKKKK} \mbox{for } \quad \| \sqrt{e^{f}-1}K_{\Lambda}\sqrt{e^{f}-1}\| < 1 & \\
         \infty  & \hphantom{KKKKK} \\
          \hphantom{KKKKKKKKKKK}\mbox{for } \quad \| \sqrt{e^{f}-1}K_{\Lambda}\sqrt{e^{f}-1}\|
            \geqslant 1,  &
        \end{cases}
\\
{\rm{(3)}} \hskip0.5cm  \int_{Q(R)}e^{i\langle f, \xi \rangle} \mu_K^{(det)}(d\xi)
   &=& \Det[1+(1-e^{if})K_{\Lambda}]^{-1},
\\
{\rm{(4)}} \ \hskip0.5cm \int_{Q(R)}e^{\langle f, \xi \rangle} \mu_K^{(det)}(d\xi)
   &=& \begin{cases}
        \Det[1-\sqrt{e^{f}-1}K_{\Lambda}\sqrt{e^{f}-1} ]^{-1} < \infty  & \hphantom{KKKKK} \\
          \hphantom{KKKKKKKKKKK}\mbox{for } \quad
      \|\sqrt{e^{f}-1}K_{\Lambda}\sqrt{e^{f}-1}\|  <1           \\
        \infty & \hphantom{KKKKK} \\
          \hphantom{KKKKKKKKKKK}\mbox{for } \quad
      \|\sqrt{e^{f}-1}K_{\Lambda}\sqrt{e^{f}-1}\|  \geqslant 1.
     \end{cases}
\end{eqnarray*}
\label{thmexp}
\end{thm}
{\sl Proof }: Let $ f \not\equiv 0 $, i.e., $\lambda (\, $supp$\, f) > 0$.
In \cite{2}, pp.213--214, it was introduced a family of symmetric non-negative functions
$\{\sigma _{\Lambda^n}\}_{n\geq0}$ defined by the equations:
\[
       \exp\big[-\rho\langle \sqrt{1-e^{-f}}, \, (1+K_f)^{-1}\sqrt{1-e^{-f}} \, \rangle \big]
    = \exp\big[-\rho\langle \chi_{\Lambda}, \, (1+(1-e^{-f})K_{\Lambda})^{-1}(1-e^{-f})
    \rangle\big]
\]
\[
  = \exp\big[-\rho \langle \chi_{\Lambda}, (1+K_{\Lambda})^{-1}\chi_{\Lambda}\rangle
          + \rho \sum_{l=0}^{\infty}\langle(1+K_{\Lambda})^{-1}\chi_{\Lambda},
           e^{-f}(R_{\Lambda}e^{-f})^l(1+K_{\Lambda})^{-1}\chi_{\Lambda}\rangle\big]
\]
\begin{equation}
       = \sum_{n=0}^{\infty}\frac{1}{n!}
         \int_{\Lambda^n}\sigma_{\Lambda^n}(x_1, \cdots, x_n)
        e^{-\sum_{k=1}^n f(x_k)} \lambda^{\otimes n}(dx_1\cdots dx_n ).
\label{2134}
\end{equation}
Here, $R_{\Lambda} = K_{\Lambda}(1+K_{\Lambda})^{-1}$ satisfies $ \|R_{\Lambda}\| <1 $
since $K_{\Lambda}$ is a bounded non-negative operator.
%For the details, we refer to Ref.\cite{2}.
Using $\{\sigma _{\Lambda^n}\}_{n\geq0}$, the random point processes $\mu_{K,\rho}$ was defined
as the probability measure such that
\begin{equation}
    \int_{Q(R)}F(\xi)\mu_{K,\rho}(d\xi) = \sum_{n=0}^{\infty}\int_{\Lambda^n}
     \sigma_{\Lambda^n}(x_1, \cdots, x_n)F(\sum_{j=1}^n\delta_{x_j})
     \lambda^{\otimes n}(dx_1, \cdots, dx_n)
\label{sig2mu}
\end{equation}
holds for any bounded (or non-negative) measurable functional satisfying
$F(\xi)=F(\xi_{\Lambda})$,  where $\xi_{\Lambda}(A) = \xi(A\cap \Lambda)$.

From this construction, we obtain the first claim (1):
\[
    \int_{Q(R)}e^{i\langle f, \xi\rangle }\mu_{K,\rho}(d\xi) =
     \sum_{n=0}^{\infty}\int_{\Lambda^n}
     \sigma_{\Lambda^n}(x_1, \cdots, x_n)e^{i\sum_{j=1}^nf(x_j)}
     \lambda^{\otimes n}(dx_1, \cdots, dx_n)
\]
\[
  = \exp[-\rho \langle \chi_{\Lambda}, (1+K_{\Lambda})^{-1}\chi_{\Lambda}\rangle
          + \rho \sum_{l=0}^{\infty}\langle(1+K_{\Lambda})^{-1}\chi_{\Lambda},
           e^{if}(R_{\Lambda}e^{if})^l(1+K_{\Lambda})^{-1}\chi_{\Lambda}\rangle]
\]
\[
      = \exp[-\rho\langle \chi_{\Lambda}, \, (1+(1-e^{if})K_{\Lambda})^{-1}(1-e^{if})
    \rangle].
\]

If $z \in \C$ satisfies $|z|e^{\|f\|_{\infty}} \leqslant 1$, then we get the equality:
\[
            \sum_{n=0}^{\infty}\int_{\Lambda^n}
     \sigma_{\Lambda^n}(x_1, \cdots, x_n)z^ne^{\sum_{j=1}^nf(x_j)}
     \lambda^{\otimes n}(dx_1, \cdots, dx_n)
\]
\begin{equation}
  = \exp[-\rho \langle \chi_{\Lambda}, (1+K_{\Lambda})^{-1}\chi_{\Lambda}\rangle
          + \rho \sum_{l=0}^{\infty}z^{l+1}\langle(1+K_{\Lambda})^{-1}\chi_{\Lambda},
           e^{f}(R_{\Lambda}e^{f})^l(1+K_{\Lambda})^{-1}\chi_{\Lambda}\rangle],
\label{+sig2exp}
\end{equation}
Since in the both sides all coefficients the $z$-power series are non-negative,
this equality (\ref{+sig2exp}) also holds for $z=1$ in the sense that either the both sides
are finite and equal or they are both diverge to $+\infty$. When they are finite, we obtain
\[
      \int_{Q(R)}e^{\langle f, \xi\rangle }\mu_{K,\rho}(d\xi) =
     \exp[\rho\langle \sqrt{e^{f}-1}, \, (1- \sqrt{e^{f}-1}K_{\Lambda}\sqrt{e^{f}-1})^{-1}
           \sqrt{e^{f}-1} \, \rangle ] \ ,
\]
cf. the proof of Theorem 2.1 in \cite{2}, pp.213-214. Hence, for the second claim (2) it is sufficient
to show that
\begin{equation}\label{equiv1-2}
    \mbox{the finite RHS of (\ref{+sig2exp})} \Leftrightarrow
    \|e^{f/2}R_{\Lambda}e^{f/2}\| < 1 \Leftrightarrow
     \|\sqrt{e^f -1}K_{\Lambda}\sqrt{e^f-1}\| < 1 \ .
\end{equation}
Notice that by Proposition 2.3(ii) \cite{2} the Condition {\bf K'}(ii)
ensures: $ R_{\Lambda}(x,y) > 0$, for $\lambda\otimes\lambda$-almost all $(x,y) \in \Lambda^2$.
Since $R_{\Lambda}$ is a compact symmetric operator, it follows from the variational
principle that $\|e^{f/2}R_{\Lambda}e^{f/2}\| $ is the largest eigenvalue of the operator
$e^{f/2}R_{\Lambda}e^{f/2} $ with eigenfunction $\varphi_0 >0 \; (\lambda\!-\!a.e. \mbox{ on } \Lambda)$.
Hence we have
\[
    \langle(1+K_{\Lambda})^{-1}\chi_{\Lambda},
           e^{f}(R_{\Lambda}e^{f})^l(1+K_{\Lambda})^{-1}\chi_{\Lambda}\rangle
\]
\[
    = |\langle\varphi_0, e^{f/2}(1+K_{\Lambda})^{-1}\chi_{\Lambda}\rangle|^2 \,
          \|e^{f/2}R_{\Lambda}e^{f/2}\|^l + O(\|e^{f/2}R_{\Lambda}e^{f/2}\|^l\delta^l)
\]
for some $\delta \in (0, 1) $.
Note that $|\langle\varphi_0, e^{f/2}(1+K_{\Lambda})^{-1}\chi_{\Lambda}\rangle| > 0$
because $(1+K_{\Lambda})^{-1}\chi_{\Lambda}>0 \; (\lambda\!-\!a.e. \mbox{ on } \Lambda$) and
 $\| (1+K_{\Lambda})^{-1}\chi_{\Lambda} \| > 0$.
Thus, we get the first equivalence in (\ref{equiv1-2}).

For the second equivalence, it is enough to prove that
\[
     \|R_{\Lambda}^{1/2}e^fR_{\Lambda}^{1/2}\| <1 \Longleftrightarrow
          \| K_{\Lambda}^{1/2}(e^f-1)K_{\Lambda}^{1/2}\| < 1
\]
by duality. Let $ \| R_{\Lambda}^{1/2}e^fR_{\Lambda}^{1/2} \| = \eta < 1$.
Then $K_{\Lambda} \geqslant 0$, $f \geqslant 0$ and
\[
     1- K_{\Lambda}^{1/2}(e^f-1)K_{\Lambda}^{1/2} = (1+K_{\Lambda})^{1/2}
         (1-R_{\Lambda}^{1/2}e^fR_{\Lambda}^{1/2})(1+K_{\Lambda})^{1/2} \ ,
\]
together with $ R_{\Lambda}^{1/2}e^fR_{\Lambda}^{1/2} \geq 0$, imply
\[
     1- K_{\Lambda}^{1/2}(e^f-1)K_{\Lambda}^{1/2}
            \geqslant (1+K_{\Lambda})(1-\eta) \geqslant 1-\eta \ .
\]
Hence $K_{\Lambda}^{1/2}(e^f-1)K_{\Lambda}^{1/2} \leqslant \eta <1$.
On the other hand, if $ \| K_{\Lambda}^{1/2}(e^f-1)K_{\Lambda}^{1/2} \|= \theta <1 $, then
\[
    1- R_{\Lambda}^{1/2}e^fR_{\Lambda}^{1/2} =
       (1+K_{\Lambda})^{-1/2}(1- K_{\Lambda}^{1/2}(e^f-1)K_{\Lambda}^{1/2})
         (1+K_{\Lambda})^{-1/2}
\]
\[
         \geqslant (1-\theta)(1+K_{\Lambda})^{-1}
      \geqslant \frac{1-\theta}{1+\|K_{\Lambda}\|} \ ,
\]
which yields
\[
        0\leqslant R_{\Lambda}^{1/2}e^fR_{\Lambda}^{1/2} \leqslant 1-
        \frac{1-\theta}{1+\|K_{\Lambda}\|}
         < 1.
\]
This finishes the proof of claims (1) and (2) of the Theorem concerning the measure $\mu_{K,\rho}$.

The claims (3) and (4) concerning the measure $\mu_K^{(det)}$ can be shown similarly if one uses,
instead of (\ref{2134}), the representation:
\[
       \Det[1+K_{f}]^{-1} = \Det[1+K_{\Lambda}]^{-1}\Det[1-e^{-f}R_{\Lambda}]^{-1}
\]
\[
      = \Det[1+K_{\Lambda}]^{-1}\sum_{n=0}^{\infty}\frac{1}{n!}\int_{R^n}
        {\rm{Per}}\{R_{\Lambda}(x_j,x_k)\}_{1\leqslant j, k\leqslant n}
    e^{-\sum_{l=1}^nf(x_l)}\lambda^{\otimes n}(dx_1, \cdots, dx_n) \ ,
\]
where $\Det$ is the \textit{Fredholm determinant} and Per is the \textit{permanent} of the
corresponding matrices \cite{ST03}.  \hfill$\square$

%%%%%%%%%%%%%%%%%%%%%%%%%%%%%%%  Operators  %%%%%%%%%%%%%%%%%%%%%%%%%%%%%%%%%%%%%%%%%%%%%%%%%
\section{Operators}
%%%%%%%%%%%%%%%%%%%%%%%%%%%%%%%%%%%%%%%%%%%%%%%%%%%%%%%%%%%%%%%%%%%%%%%%%%%%%%%%%%%%%%%%%%
Below we deal with the boson random point processes which describe the position distribution
of the perfect Bose-gas ($\R^d$ for $d > 2$) \textit{above} the critical particle density
$\rho_{c} := \rho_{c}(\beta)$, i.e. in the regime of the Bose-Einstein condensation.

To this end we set $ R := \R^d$ and $K^{\beta} := G^{\beta}(1-G^{\beta})^{-1}$ for $K$, where
$G^{\beta} := e^{\beta\Delta}$ for $G$. Here $\beta>0$ is the inverse temperature and
$\Delta$ denotes
the $d$-dimensional self-adjoint Laplacian operator in the space $L^2(\R^d)$ equipped by the
Lebesgue measure.
Then it can be shown that operator $K^{\beta}$ satisfies the Condition {\bf K'}, see \cite{2}.

In the present section, we derive some miscellaneous properties of the operators, which
we use in the
line of reasoning of the next section. First we adopt the following definition of the
Fourier transformation:
\[
    \widetilde h(p) := (\mathscr{F}h)(p) =
            \int_{\R^d}e^{-ip\cdot x}h(x)\frac{dx}{(2\pi)^{d/2}}
\]
for $ h \in L^1(\R^d) $ and for its extension to $L^2(\R^d)$.
%%%%%%%%%%%%%%%%%%%%% LEMMA %%%%%%%%%%%%%%%%%%%%%%%%%%%%
\begin{lem}
For any compact $\Lambda \subset \R^d$ the operator
$(-\Delta)^{-1/2}\chi_{\Lambda}, (K^{\beta})^{1/2}\chi_{\Lambda}$ is bounded.  Therefore,
\[
   (-\Delta)^{-1}_{\Lambda} := \big((-\Delta)^{-1/2}\chi_{\Lambda}\big)^*
  (-\Delta)^{-1/2}\chi_{\Lambda}
\]
\[
           K^{\beta}_{\Lambda} := \big((K^{\beta})^{-1/2}\chi_{\Lambda}\big)^*
          (K^{\beta})^{-1/2}\chi_{\Lambda}
\]
are bounded non-negative self-adjoint operators.
\label{Linv}
\end{lem}
%%%%%%%%%%%%%%%%%%%%%%%%%%%%%%%%%%%%%%%%%%%%%%%%%%%%%%%%%%%%%%%%%%%%%%%%%%%%%%%%%%%%%%%%%%
{\sl Proof }: These properties can be verified with a help of the Fourier transformation.
For any $g \in L^2(\R^d)$, we obtain:
\[
     \|  (-\Delta)^{-1/2}\chi_{\Lambda}g\|^2_2 =
      \int_{\R^d} \frac{|\widetilde{\chi_{\Lambda}g}(p)|^2}{|p|^2}\, dp
\]
\[
     \leqslant \int_{|p| \leqslant 1}\frac{\|\widetilde{\chi_{\Lambda}g}\|_{\infty}^2}
     {|p|^2}\, dp + \int_{\R^d}|\widetilde{\chi_{\Lambda}g}(p)|^2\, dp
\]
\[
     \leqslant c_1 \|\chi_{\Lambda}g\|_1^2 + \|\chi_{\Lambda}g\|_2^2
     \leqslant c_2 \|\chi_{\Lambda}\|_2^2\|g\|_2^2 +\|\chi_{\Lambda}\|_{\infty}^2\|g\|_2^2
\]
\[
    =(1+c|\Lambda|)\|g\|_2^2.
\]
Thus, $(-\Delta)^{-1/2}\chi_{\Lambda}$ is bounded and
$\|(-\Delta)^{-1/2}\chi_{\Lambda}\| \leqslant \sqrt{1+c|\Lambda|}$ holds.
It gives $\|(-\Delta)^{-1}_{\Lambda}\| \leqslant 1+c|\Lambda|$.
Here, $|\Lambda|$ denotes the Lebesgue measure of $\Lambda$.
%The constants $c > 0$ vary line to line, however they depend only on $d>2$.

A similar argument is valid for the operator $K^{\beta}_{\Lambda}$. \hfill $\square$

%\medskip

\begin{definition}  For $\kappa > 0$, we define the transformation
\[
       U_{\kappa}: L^2(\R^d) \ni g(\, \cdot \,) \mapsto
         \kappa^{d/2}g( \kappa \, \cdot \,) \in L^2(\R^d).
\]
\end{definition}
%Then we have the following statement:
%%%%%%%%%%%%%%%%%%%%%%%%%%%%%%%%  LEMMA U %%%%%%%%%%%%%%%%%%%%%%%%%%%%%%%%%%%%%%%%%%%%%%%%%%%
\begin{lem}
The transformation $U_{\kappa}$ is unitary on $L^2(\R^d)$ for any $\kappa >0$, and it has
the following properties:

{\rm{(1)}} \quad $\displaystyle U_{\kappa} h\, U_{\kappa}^{-1} = h(\kappa \, \cdot \, )$ \quad for
    the multiplication operator by function $h$ \ .

{\rm{(2)}} \quad $\displaystyle U_{\kappa}  \Delta  U_{\kappa}^{-1} = \kappa^{-2} \Delta $ \ .

{\rm{(3)}} \quad $\displaystyle U_{\kappa}(-\Delta)^{-1}_{\kappa\Lambda} U_{\kappa}^{-1} =
     \kappa^2(-\Delta)^{-1}_{\Lambda}, \qquad
     U_{\kappa} G^{\beta} U_{\kappa}^{-1} = G^{\beta/\kappa^{2}} $ \ .

{\rm{(4)}} \quad $\displaystyle U_{\kappa} K^{\beta}_{\kappa\Lambda} U_{\kappa}^{-1}
         = K^{\beta/\kappa^{2}}_{\Lambda}$ \ .

\label{U}
\end{lem}
%%%%%%%%%%%%%%%%%%%%%%%%%%%%%%%%%%%%%%%%%%%%%%%%%%%%%%%%%%%%%%%%%%%%%%%%%%%%%%%%%%%%%%%%%%%%%%%%%%%%
{\sl Proof }:  These properties are a straightforward consequence of the
relation $\mathscr{F}U_{\kappa} = U_{\kappa}^{-1}\mathscr{F}$
on $L^2(\R^d)$.  \hfill $\square$

%\medskip

%%%%%%%%%%%%%%%%%%%%%%%%%%%%%%%%%%%% DEF %%%%%%%%%%%%%%%%%%%%%%%%%%%%%%%%%%%%%%%%%%%%%%%%%%%%%%%%%%%
\begin{definition}\label{defPM}
For bounded non-negative function $f$ with a compact support
and for $\kappa > 0$, we put
\[
      f_{\kappa}^{(\pm)}(x) := \pm \kappa^2\big(e^{\pm f(x)/\kappa^2} -1\big) \ .
\]
\end{definition}

%The following bounds hold:
%%%%%%%%%%%%%%%%%%%%%%%%%%%%% LEMMA fmp %%%%%%%%%%%%%%%%%%%%%%%%%%%%
\begin{lem}
One has the following estimates:
\[
      f^{(\pm)}_{\kappa}(x) \geqslant 0, \quad
    \bigg| \chi_{\{f > 0\}}(x)\frac{f_{\kappa}^{(\pm)}(x)}{f(x)}\bigg|
     \leqslant e^{\|f\|_{\infty}/\kappa^2}, \quad
     \chi_{\{f > 0\}}(x)\bigg| 1- \sqrt{\frac{f_{\kappa}^{(\pm)}(x)}{f(x)}}\bigg|
     \leqslant \frac{\|f\|_{\infty}}{2\kappa^2} e^{\|f\|_{\infty}/\kappa^2},
\]
\[
   \|f_{\kappa}^{(\pm)}\|_{\infty} \leqslant \|f\|_{\infty}e^{\|f\|_{\infty}/\kappa^2}, \quad
      \|f-f_{\kappa}^{(\pm)}\|_{\infty} \leqslant
              \frac{\|f\|_{\infty}}{2\kappa^2} e^{\|f\|_{\infty}/\kappa^2}.
\]
\label{fmp}
\end{lem}
%%%%%%%%%%%%%%%%%%%%%%%%%%%%%%%%%%%%%%%%%%%%%%%%%%%%%%%%%%%%%%%%%%%%%%
{\sl Proof }: These estimates are a direct consequence of the elementary inequalities:
\[
        \frac{|e^y-1|}{|y|} \leqslant e^{|y|}, \quad
\frac{|e^y-1-y|}{|y|} \leqslant
       \frac{|y|e^{|y|}}{2} \  \ ,
\]
and $|\sqrt z-1|\leqslant |z-1|$ for $ y \in \R-\{0\},
         z \geqslant 0 $.  \hfill $\square$

%\medskip

%From those preliminaries, we obtain the following estimates for the operators.

%%%%%%%%%%%%%%%%%%%%%%%%%%%%%%%%%%%%%% LEMMA K2L %%%%%%%%%%%%%%%%%%%%%%
\begin{lem} For any $\kappa > 0$ we have the estimates:
\[
    0 \leqslant (-\beta\Delta)_{\Lambda}^{-1} -
        \kappa^{-2}K_{\Lambda}^{\beta/\kappa^{2}} \leqslant (2\kappa^2)^{-1} \ \ .
\]
\label{K2L}
\end{lem}
%%%%%%%%%%%%%%%%%%%%%%%%%%%%%%%%%%%%%%%%%%%%%%%%%%%%%%%%%%%%%%%%%%%%%%%%%%%
{\sl Proof }: Using the Fourier transformation, we get
\[
       \langle g,\big[\kappa^2(-\beta\Delta)_{\Lambda}^{-1}
      - K_{\Lambda}^{\beta/\kappa^{2}}\big]g\rangle
       = \int_{\R^d}\Big(\frac{\kappa^2}{\beta|p|^2}
         - \frac{1}{e^{\beta|p|^2/\kappa^2}-1}\Big)|\widetilde{\chi_{\Lambda}g}(p)|^2\, dp.
\]
Then lemma follows from the inequality
\[
         0 \leqslant \frac{1}{y} - \frac{1}{e^y-1} \leqslant \frac{1}{2} \ \ ,
 \qquad \mbox{for } \; y> 0 \ ,
\]
and from  the estimate $\|\widetilde{\chi_{\Lambda}g}\|_2 = \|\chi_{\Lambda}g\|_2 \leqslant \|g\|_2$ \ .
\hfill $\square$

%%%%%%%%%%%%%%%%%%%%%%%%%%%%%%% LEMMA K2Lf %%%%%%%%%%%%%%%%%%%%%%%%%%%%%%%%%
\begin{lem}
Suppose that $\supp f \subset \Lambda $. Then for $\kappa \rightarrow 0$ one gets the
operator-norm asymptotics:
\[
     \| \sqrt f(-\beta\Delta)_{\Lambda}^{-1}\sqrt f
   - \kappa^{-2} \sqrt{f_{\kappa}^{(\pm)}}K^{\beta/\kappa^2}_{\Lambda}\sqrt{f_{\kappa}^{(\pm)}}
    \| = O(\kappa^{-2}) \ ,
\]
in the space $L^2(\R^d)$.

\label{K2Lf}
\end{lem}
%%%%%%%%%%%%%%%%%%%%%%%%%%%%%%%%%%%%%%%%%%%%%%%%%%%%%%%%%%%%%%%%%%%%%%%%%%%
{\sl Proof }: From Lemma \ref{Linv}, \ref{fmp} and \ref{K2L}, we obtain
\[
    \| \sqrt f(-\beta\Delta)_{\Lambda}^{-1}\sqrt f
   - \kappa^{-2} \sqrt{f_{\kappa}^{(\pm)}}K^{\beta/\kappa^2}_{\Lambda}\sqrt{f_{\kappa}^{(\pm)}}
   \|
\]
\vspace{-10mm}
\begin{eqnarray*}
 & \leqslant &   \| ( \sqrt f - \sqrt{f_{\kappa}^{(\pm)}})
 (-\beta\Delta)_{\Lambda}^{-1} \sqrt f \|
+ \| \sqrt{f_{\kappa}^{(\pm)}} (-\beta\Delta)_{\Lambda}^{-1}
   ( \sqrt{f} - \sqrt{f_{\kappa}^{(\pm)}})\|
\\
 & + & \|\sqrt{f_{\kappa}^{(\pm)}}[ (-\beta\Delta)_{\Lambda}^{-1}
   - \kappa^{-2}K^{\beta/\kappa^2}_{\Lambda}]\sqrt{f_{\kappa}^{(\pm)}}
    \|
\\
  & \leqslant & (\|\sqrt{f}\|_{\infty} +\| \sqrt{f_{\kappa}^{(\pm)}}\|_{\infty})
       \| (-\beta\Delta)_{\Lambda}^{-1}\| \,
       \| \sqrt{f}- \sqrt{f_{\kappa}^{(\pm)}} \|_{\infty}
\\
  & + & \|\sqrt{f_{\kappa}^{(\pm)}}\|^2_{\infty}
    \| (-\beta\Delta)_{\Lambda}^{-1}  - \kappa^{-2}K^{\beta/\kappa^2}_{\Lambda}\|
 = O(\kappa^{-2}) \ . \qquad \qquad \qquad  \qquad \qquad \qquad \square
\end{eqnarray*}
%%%%%%%%%%%%%%%%%%%%%%%%%%%%% LEMMA KTR %%%%%%%%%%%%%%%%%%%%%%%%%%%%%%%%
\begin{lem}
The operator $K_{\Lambda}^{\beta/\kappa^2}\in\mathfrak{C}_{1}(L^2(\R^d))$, i.e. belongs to
the trace-class
operators on $L^2(\R^d)$, and
\begin{equation}
     \Tr [\sqrt f K_{\Lambda}^{\beta/\kappa^2} \sqrt f]  =
         \kappa^d\rho_c \int_{\R^d} f(x) \, dx \ .
\label{KfTR}
\end{equation}
\label{KTR}
\end{lem}
%%%%%%%%%%%%%%%%%%%%%%%%%%%%%%%%%%%%%%%%%%%%%%%%%%%%%%%%%%%%%%%%%%%%%%%%%%%%%%%%%%%%%%%%%%
{\sl Proof }:
Let $\{ \, \phi_n \, \}_{n}$ be a \textit{complete ortho-normal system} (CONS)
functions in $L^2(\R^d)$ and let
$g(x) =e^{ip\cdot x}\chi_{\Lambda}(x)$. Then we have
\[
       \sum_n|\widetilde{\chi_{\Lambda}\phi_n} (p)|^2 =
      \sum_n\frac{|\langle g, \phi_n\rangle|^2}{(2\pi)^d} = \frac{\|g\|_2^2}{(2\pi)^d}
      = \frac{\|\chi_{\Lambda}\|_2^2}{(2\pi)^d} \ .
\]
This yields
\[
   \sum_n\langle \phi_n, K^{\beta/\kappa^2}_{\Lambda} \phi_n\rangle
  = \sum_n \int_{\R^d}\frac{1}{e^{\beta|p|^2/\kappa^2}-1}
               |\widetilde{\chi_{\Lambda}\phi_n} (p)|^2 dp
\]
\[
     = \|\chi_{\Lambda}\|^2_2 \int_{\R^d}\frac{1}{e^{\beta|p|^2/\kappa^2}-1}
               \frac{dp}{(2\pi)^d} = \kappa^d\rho_c|\Lambda | < \infty \ .
\]
Since $K_{\Lambda}^{\beta/\kappa^2} \geq 0$, it follows that
$K_{\Lambda}^{\beta/\kappa^2}\in\mathfrak{C}_{1}(L^2(\R^d))$.
Similarly, we obtain the explicit value (\ref{KfTR}).
\hfill $\square$

%%%%%%%%%%%%%%%%%%%%%%%% LEMMA KHS %%%%%%%%%%%%%%%%%%%%%%%%%%%%%%%%%%%%%%%%%%%%%%%%%%%%%
\begin{lem} The operator $K_{\Lambda}^{\beta/\kappa^2} \geq 0$ verifies the following
Hilbert-Schmidt norm estimate from above:
    \begin{equation}\label{est-HS}
     \| K_{\Lambda}^{\beta/\kappa^2}\|^2_{HS} \leqslant
      c_d \big( \kappa^2/\beta\big)^{(d\vee 4)/2}\big(1+|\log (\kappa^2/\beta)| \big)
          |\Lambda| (1+ |\Lambda|) \ .
    \end{equation}
Here $c_d$ is a constant depending only on the dimension $d>2$.
\label{KHS}
\end{lem}
%%%%%%%%%%%%%%%%%%%%%%%%%%%%%%%%%%%%%%%%%%%%%%%%%%%%%%%%%%%%%%%%%%%%%%%%%%%%%%%%%%%%%%%%%
{\sl Proof: } By the Fourier transformation, we obtain
\begin{eqnarray}
   \| K^{\beta/\kappa^2}_{\Lambda} \|_{HS}^2 &=& \int_{\R^d}\frac{dq}{(2\pi)^{d/2}}
\int_{\R^d}\frac{dp}{(2\pi)^{d/2}}
   \frac{|\widetilde{ \chi_{\Lambda}}(p-q)|^2}{(e^{\beta|p|^2/\kappa^2}-1)
(e^{\beta|q|^2/\kappa^2}-1)}
\label{KHS1} \\
   &=& \int_{\R^d}\frac{dq}{(2\pi)^{d/2}}\int_{\R^d}\frac{dp}{(2\pi)^{d/2}}
    \frac{|\widetilde{ \chi_{\Lambda}}(p)|^2}
     {(e^{\beta|p+q|^2/\kappa^2}-1)(e^{\beta|q|^2/\kappa^2}-1)} \ .
\label{KHS2}
\end{eqnarray}
$1^{\circ}. { \ \rm{Case:}} \quad 2<d<4 $.

From (\ref{KHS2}), we obtain the estimate:
\begin{eqnarray*}
    \| K^{\beta/\kappa^2}_{\Lambda} \|_{HS}^2 &\leqslant&
    \int_{\R^d}\frac{dq}{(2\pi)^{d/2}}\int_{\R^d}\frac{dp}{(2\pi)^{d/2}}
 \frac{\kappa^4|\widetilde{ \chi_{\Lambda}}(p)|^2}
     {\beta^2|p+q|^2|q|^2}
\\
    &=& \bigg(\frac{\kappa^2}{\beta}\bigg)^{4/2}
      \int_{\R^d}  \frac{|\widetilde{\chi_{\Lambda}}(p)|^2}{|p|^{4-d}}\, \frac{dp}{(2\pi)^{d}}
    \int_{\R^d} \frac{d\tilde q}{|e+\tilde q|^2|\tilde q|^2}
\\
    &\leqslant& \bigg(\frac{\kappa^2}{\beta}\bigg)^{4/2}
     c \bigg[\int_{|p|\leqslant 1} \frac{\|\widetilde{ \chi_{\Lambda}}\|_{\infty}^2}
      {|p|^{4-d}}\, dp  + \int_{|p|>1} |\widetilde{\chi_{\Lambda}}(p)|^2 \, dp\bigg]
\\
   &\leqslant& \bigg(\frac{\kappa^2}{\beta}\bigg)^{4/2}
     c\big( \|\chi_{\Lambda}\|_1^2 + \|\chi_{\Lambda}\|_2^2\big)
    = \bigg(\frac{\kappa^2}{\beta^2}\bigg)^{4/2}
     c(|\Lambda|^2 + |\Lambda|) \ .
\end{eqnarray*}
Here we changed the variable $ q=|p|\tilde q $  in the
first equality, and we denote by $e$ a unit vector in $\R^d$.

\noindent $ 2^{\circ}. { \ \rm{Case:}} \quad d>4 $.

We apply the Cauchy-Schwarz inequality to (\ref{KHS1}) to get
\begin{eqnarray*}
   \| K^{\beta/\kappa^2}_{\Lambda} \|_{HS}^2 &\leqslant&
  \sqrt{\int\int
   \frac{|\widetilde{ \chi_{\Lambda}}(p-q)|^2\, dpdq}{(e^{\beta|p|^2/\kappa^2}-1)^2(2\pi)^{d}}}
   \sqrt{\int\int\frac{|\widetilde{ \chi_{\Lambda}}(p-q)|^2\, dpdq}
      {(e^{\beta|q|^2/\kappa^2}-1)^2(2\pi)^d}}
\\
   &=& \int_{\R^d}|\widetilde{ \chi_{\Lambda}}(p)|^2\, \frac{dp}{(2\pi)^{d}}\int_{\R^d}
        \frac{dq}{(e^{\beta|q|^2/\kappa^2}-1)^2}
\\
    &=&   c_d|\Lambda|\bigg(\frac{\kappa^2}{\beta}\bigg)^{d/2}.
\end{eqnarray*}

\noindent $3^{\circ}. { \ \rm{Case:}} \quad d=4 $.

Let us decompose (\ref{KHS2}) into two parts:
\begin{eqnarray*}
    \| K^{\beta/\kappa^2}_{\Lambda} \|_{HS}^2 &=&
    \int_{\R^d}\frac{dp}{(2\pi)^{d}} \, |\widetilde{ \chi_{\Lambda}}(p)|^2
    \bigg\{\int_{|q|\geqslant 2|p|} + \int_{|q|< 2|p|}\bigg\}\frac{dq}
     {(e^{\beta|p+q|^2/\kappa^2}-1)(e^{\beta|q|^2/\kappa^2}-1)}
\\
    &=& {\cal I}_1 + {\cal I}_2.
\end{eqnarray*}

For ${\cal I}_1$, $|q|\geqslant 2|p|$ implies $|p+q| \geqslant |q|-|p| \geqslant |q|/2$.
Therefore, it follows that
\begin{eqnarray*}
   {\cal I}_1 &\leqslant& \int_{\R^d}\frac{dp}{(2\pi)^{d}} \,
|\widetilde{ \chi_{\Lambda}}(p)|^2
    \int_{|q|\geqslant 2|p|}\frac{dq}
     {(e^{\beta|q|^2/4\kappa^2}-1)(e^{\beta|q|^2/\kappa^2}-1)}
\\
    &\leqslant& \int_{\R^d}\frac{dp}{(2\pi)^{d}} \, |\widetilde{ \chi_{\Lambda}}(p)|^2
    \bigg\{\theta(1-2|p|)\int_{1\geqslant |q|\geqslant 2|p|}
     + \int_{\kappa/\sqrt{\beta} \geqslant |q| > 1} + \int_{|q|> \kappa/\sqrt{\beta}}\bigg\}
\\
   && \qquad \times \frac{dq}
     {(e^{\beta|q|^2/4\kappa^2}-1)(e^{\beta|q|^2/\kappa^2}-1)}
\end{eqnarray*}
\begin{eqnarray*}
   &\leqslant& \int_{\R^d}\frac{dp}{(2\pi)^{d}} \, |\widetilde{ \chi_{\Lambda}}(p)|^2
    \bigg\{\theta(1-2|p|)\int_{1\geqslant |q|\geqslant 2|p|}
    \frac{4\kappa^4 \, dq}{\beta^2 \, |q|^4}
     + \int_{\kappa/\sqrt{\beta} \geqslant |q| > 1}
      \frac{4\kappa^4 \, dq}{\beta^2 \, |q|^4}
\\
   && \quad  + \int_{|q|> \kappa/\sqrt{\beta}}
    \frac{dq}
     {(e^{\beta|q|^2/4\kappa^2}-1)(e^{\beta|q|^2/\kappa^2}-1)}\bigg\}
\\  &\leqslant& \int_{\R^d}\frac{dp}{(2\pi)^{d}} \, |\widetilde{ \chi_{\Lambda}}(p)|^2
    \bigg\{\theta(1-2|p|)\frac{\kappa^4 }{\beta^2 } c_1 \log \frac{1}{2|p|}
     + \frac{\kappa^4 }{\beta^2} c_2\Big|\log\Big(\frac{\kappa^2 }{\beta}\Big)\Big|
\\
   && \quad  + \Big(\frac{\kappa^2 }{\beta}\Big)^{2}\int_{|\tilde q|> 1}
    \frac{d\tilde q}
     {(e^{|\tilde q|^2/4}-1)(e^{|\tilde q|^2}-1)}\bigg\}
\\
   &\leqslant& \|\widetilde{\chi_{\Lambda}}\|_{\infty}^2\Big(\frac{\kappa^2}{\beta}\Big)^{4/2}
    c_1\int_{|p|\leqslant 1/2}\log \frac{1}{2|p|} \, dp +
    \Big(\frac{\kappa^2 }{\beta}\Big)^{4/2}
   \Big(c_2\Big|\log\Big(\frac{\kappa^2 }{\beta}\Big)\Big|
      +c_3\Big)\|\chi_{\Lambda}\|^2_2
\\
    &\leqslant& c_4 \, \Big(\frac{\kappa^2}{\beta}\Big)^{4/2}
\big(1+\big|\log(\kappa^2/\beta)\big|\big)
    \big(|\Lambda| + |\Lambda|^2\big) \ .
\end{eqnarray*}

For ${\cal I}_2$, we obtain:
\begin{eqnarray*}
    {\cal I}_2 &\leqslant & \int_{\R^4}\frac{dp}{(2\pi)^{d}}
         |\widetilde{\chi_{\Lambda}}(p)|^2 \int_{|q|<2|p|}
    \frac{\kappa^4 dq} {\beta^2|p+q|^2|q|^2}
\\
     &= & \int_{\R^d}\frac{dp}{(2\pi)^{d}} |\widetilde{\chi_{\Lambda}}(p)|^2
    \int_{|\tilde q| <2}
    \frac{\kappa^4 d\tilde q}{\beta^2|e+ \tilde q|^2|\tilde q|^2}
\\
      &=& c \, \Big( \frac{\, \kappa^2}{\beta \; } \Big)^{4/2} |\Lambda|.
\end{eqnarray*}
Thus, we have obtained the desired estimate (\ref{est-HS}) for all cases. \hfill $\square$

%%%%%%%%%%%%%%%%%%%%%%%%%%%%%% Limit THMs for BEC %%%%%%%%%%%%%%%%%%%%%%%%%
\section{Limit theorems for BEC}
%%%%%%%%%%%%%%%%%%%%%%%%%%%%%%%%%%%%%%%%%%%%%%%%%%%%%%%%%%%%%%%%%%%%%%%%%%%%
In this section, we consider the boson random point processus (perfect Bose-gas) in
the \textit{regime condensation}, i.e.  when
\[
     \rho > \rho_c (= \rho_c (\beta)) \quad \mbox{ and } \quad
    \nu_{\rho} = \mu_{K^{\beta}}^{(det)} \ast \mu_{K^{\beta},(\rho-\rho_c)} \ ,
\]
where $\displaystyle \rho_c (\beta)= K^{\beta}(x,x) =
\int_{\R^d}\frac{1}{e^{\beta|p|^2}-1}\frac{dp}{(2\pi)^d}$.

%%%%%%%%%%%%%%%%%%%%%%%% Prop MV %%%%%%%%%%%%%%%%%%%%%%%%%%%
\begin{prop} For a bounded measurable set $\Lambda \subset \R^d$ and
non-negative bounded function $f$ with $\supp f \subset \Lambda$, one gets the equalities:
\[
   \int_{Q(\R^d)}\langle f, \xi \rangle \nu_{\rho}(d\xi) = \rho\int_{\R^d}f(x)\,dx \ ,
\]
and
\[
   \int_{Q(\R^d)}\Big( \langle f, \xi \rangle - \int_{Q(\R^d)}\langle f, \xi \rangle
    \nu_{\rho}(d\xi)\Big)^2 \nu_{\rho}(d\xi) = \rho\int_{\R^d}f(x)^2\,dx
             + \Tr[fK_{\Lambda}^{\beta}fK_{\Lambda}^{\beta}] + 2(\rho - \rho_c)
          \langle f, K_{\Lambda}^{\beta}f \rangle.
\]
\label{MV}
\end{prop}
%%%%%%%%%%%%%%%%%%%%%%%%%%%%%%%%%%%%%%%%%%%%%%%%%%%%%%%%%%%%%%%
{\sl Proof }: Let us put
\[
     e^{-W(f)} := \int_{Q(\R^d)}e^{-\langle f, \xi \rangle} \nu_{\rho}(d\xi) \ ,
\]
then from (\ref{chdet}) and (\ref{chrho}) we get
\[
    W(f) = (\rho - \rho_c)\langle \chi_{\Lambda},
      (1+(1-e^{-f})K_{\Lambda}^{\beta})^{-1}(1-e^{-f})\rangle
      +\log \Det[1+(1-e^{-f})K_{\Lambda}^{\beta}] \ .
\]
For small $\epsilon >0$, this yields the expansion:
\[
     W(\epsilon f) = \epsilon \rho\int_{\R^d}f(x)\,dx
   -\frac{\epsilon^2}{2} \rho\int_{\R^d}f(x)^2\,dx
   -\frac{\epsilon^2}{2}\Tr[fK_{\Lambda}^{\beta}fK_{\Lambda}^{\beta}]
      - \epsilon^2(\rho - \rho_c)\langle f, K_{\Lambda}^{\beta}f \rangle
         +O(\epsilon^3),
\]
which implies the proposition. \hfill $\square$

%%%%%%%%%%%%%%%%%%%%%%%%%%%%%%%%%%% Cor %%%%%%%%%%%%%%%%%%%%%%%%%%%%
\begin{corollary} Under the same conditions as in the Proposition 4.1, one obtains, for large $\kappa$,
the following asymptotics :
\[
    \int_{Q(\R^d)}\Big\langle f\Big(\frac{.}{\kappa}\Big), \xi \Big\rangle
     \nu_{\rho}(d\xi) = \kappa^d\rho\int_{\R^d}f(x)\,dx  + {o}(\kappa^{d}) \ ,
\]
and
\begin{eqnarray*}
&&\int_{Q(\R^d)}\Big( \Big\langle f\Big(\frac{.}{\kappa}\Big), \xi \Big\rangle
- \int_{Q(\R^d)}\Big\langle f\Big(\frac{.}{\kappa}\Big), \xi \Big\rangle
\nu_{\rho}(d\xi)\Big)^2 \nu_{\rho}(d\xi)  \nonumber \\
&&=  2\kappa^{d+2}(\rho - \rho_c)
\langle f, ( -\beta{\Delta})^{-1}_{\Lambda}f \rangle + O(\kappa^{4\vee d}\log\kappa) \ .
\end{eqnarray*}
\label{sMV}
\end{corollary}
%%%%%%%%%%%%%%%%%%%%%%%%%%%%%%%%%%%%%%%%%%%%%%%%%%%%%%%%%%%%%%%%%%%%%%
{\sl Proof }: Using the unitary operator $U_{\kappa}$, we get
\[
    \Tr [f(\, \cdot/ \kappa)K_{\kappa\Lambda}^{\beta}
      f(\, \cdot/ \kappa)K_{\kappa\Lambda}^{\beta}]
    = \Tr [U_{\kappa}f(\, \cdot/ \kappa)K_{\kappa\Lambda}^{\beta}
       f(\, \cdot/ \kappa)K_{\kappa\Lambda}^{\beta}U_{\kappa}^{-1}]
\]
\[
      =  \Tr [fK_{\Lambda}^{\beta/\kappa^{2}}
       fK_{\Lambda}^{\beta/\kappa^{2}}]
       \leqslant \|f\|_{\infty}^2 \| K_{\Lambda}^{\beta/\kappa^{2}} \|_{HS}^2
       =O(\kappa^{d\vee 4}\log\kappa)
\]
and
\[
     \langle f( \, \cdot/\kappa), K_{\kappa\Lambda}^{\beta}f( \, \cdot/\kappa) \rangle
     =  \langle U_{\kappa}f( \, \cdot/\kappa),  U_{\kappa}
      K_{\kappa\Lambda}^{\beta}f( \, \cdot/\kappa) \rangle
\]
\[
   = \kappa^d \langle f, K_{\Lambda}^{\beta\kappa^{-2}}f \rangle
   = \kappa^{d+2} \langle f, (-\beta\Delta)_{\Lambda}^{-1}f \rangle +O(\kappa^d).
\]
Here we used Lemma \ref{KHS} and Lemma \ref{K2L}.
Note that $ \supp f( \, \cdot/\kappa) \subset \kappa\Lambda$.
Then Proposition \ref{MV} yields the Corollary. \hfill $\square$

%%%%%%%%%%%%%%%%%%%%%%%%%%%%%%%%%%%%%% THN thmL %%%%%%%%%%%%%%%%%%%%%%%%%%%%%%%%%%%%%%%%%%%%%%%%%%%%
\begin{thm}[The law of large number] \; For $\kappa \to \infty$ and for any bounded function $f$
with compact support the limit
\[
    \frac{1}{\kappa^d}\Big\langle f\Big(\frac{.}{\kappa}\Big), \xi \Big\rangle
      \longrightarrow \rho\int_{\R^d}f(x)\,dx
\]
holds in $L^2(Q(\R^d), \nu_{\rho})$.
\label{thmL}
\end{thm}
{\sl Proof }: This is a simple consequence of the Corollary \ref{sMV}. \hfill $\square$

%%%%%%%%%%%%%%%%%%%%%%%%%%%%%%%%%%%%% THM CLT %%%%%%%%%%%%%%%%%%%%%%%%%%%%%%%%%%%%%%%%%%%%%%%%%%%%%%%
\begin{thm}[Central Limit Theorem]
For $\kappa \to \infty$ the family of random variables
\[
  X_{\kappa} = \kappa^{-(d+2)/2} \ \ \frac{\langle f(\,\cdot/\kappa),\xi\rangle
         - \rho\kappa^d\int_{\R^d}f(x)\,dx}
     {\sqrt{2(\rho -\rho_c)\langle f, (-\beta\Delta)_{\Lambda}^{-1} f\rangle}}
\]
converges in distribution to the standard Gaussian random variable.
\label{CLT}
\end{thm}
%%%%%%%%%%%%%%%%%%%%%%%%%%%%%%%%%%%%%%%%%%%%%%%%%%%%%%%%%%%%%%%%%%%%%%%%%
{\sl Proof }: By Theorem \ref{thmexp}(1),(3), we obtain
\[
     \mathbb{E}_{\nu_{\rho}}\Big[ \exp\big[i\lambda\kappa^{-(d+2)/2}
       \big(\langle f(\,\cdot/\kappa),\xi\rangle - \rho\kappa^d\int_{\R^d}f(x)\,dx\big)\big]\Big]
\]
\[
        =\exp\big[-i\lambda\kappa^{(d-2)/2}\rho\int_{\R^d} f(x)\,dx-W_{\kappa}\big],
\]
where
\[
     W_{\kappa} =  (\rho-\rho_c) \langle \chi_{\kappa\Lambda},
     (1+(1-e^{i\lambda\kappa^{-(d+2)/2}f(\cdot/\kappa)})K^\beta_{\kappa\Lambda})^{-1}
     (1-e^{i\lambda\kappa^{-(d+2)/2}f(\cdot/\kappa)}) \rangle
\]
\[
         +\log\Det[1+(1-e^{i\lambda\kappa^{-(d+2)/2}f(\cdot/\kappa)})K^\beta_{\kappa\Lambda}].
\]
By definition of transformation $U_{\kappa}$ and by Lemma \ref{K2L}, the first term can be expanded as
\[
   (\rho-\rho_c) \langle U_{\kappa}\chi_{\kappa\Lambda},
     U_{\kappa}(1+(1-e^{i\lambda\kappa^{-(d+2)/2}f(\cdot/\kappa)})
           K^{\beta}_{\kappa\Lambda})^{-1}
     (1-e^{i\lambda\kappa^{-(d+2)/2}f(\cdot/\kappa)}) \rangle
\]
\[
    = -i\lambda(\rho-\rho_c)\kappa^{(d-2)/2}\Big[ \int f\,dx
       + i\lambda\kappa^{-(d+2)/2}\langle f, K^{\beta/\kappa^2}_{\Lambda} f
      \rangle\Big] +o(1)
\]
\[
     =-i\lambda(\rho-\rho_c)\kappa^{(d-2)/2}\int f\,dx
       +\lambda^2(\rho -\rho_c)\langle f, (-\beta\Delta)_{\Lambda}^{-1}f \rangle +o(1).
\]
Here we applied the bound:
\[
          \|(1-Y)^{-1} - (1 + Y) \| \leqslant c \| Y \|^2
\]
valid for operators with small enough operator norms with a bound defined by $c$.

Similarly, we get also the representation for the second term:
\[
     \log \Det \Big[ 1+(1-e^{i\lambda\kappa^{-(d+2)/2}f})K^{\beta/\kappa^2}_{\Lambda}\Big]
\]
\[
       = -i\lambda\kappa^{-(d+2)/2}\Tr \big[fK^{\beta/\kappa^2}_{\Lambda}\big] + R \ ,
\]
where
\[
         \Tr \big[fK^{\beta/\kappa^2}_{\Lambda}\big] = \rho_c\kappa^d\int f(x)\,dx \ ,
\]
and
\[
    |R| \leqslant \Tr \Big[ \big( (1-e^{i\lambda\kappa^{-(d+2)/2}f})
       K^{\beta/\kappa^2}_{\Lambda} \big)^2 \Big]
\]
\[
       =O(\lambda^2\kappa^{-d-2})\|f\|_{\infty}^2\|K^{\beta/\kappa^2}_{\Lambda}\|_{HS}^2
       =o(1) \ .
\]
Here we used the bound:
\begin{equation}
      | \log \Det[1+Y] -\Tr Y| = |\log \Det_2[1+Y]| = O(\| Y \|_{HS}^2)
\label{Det_2}
\end{equation}
for the trace-class operators with small operator norms. Recall that 
$\Det_2[1+Y]:= e^{-\Tr Y} \Det[1+Y] = \Det[(1-Y) e^{-Y}]$ denotes a "regularized"
determinant for the Hilbert-Schmidt operators $Y$, see e.g. \cite{ST03}.

Thus we get
\[
       W_{\kappa} = -i\lambda\rho\kappa^{(d-2)/2}\int f\,dx
       +\lambda^2(\rho -\rho_c)\langle f, (-\beta\Delta)_{\Lambda}^{-1}f \rangle +o(1)
\]
and
\[
     \mathbb{E}_{\nu_{\rho}}\Big[ \exp\big[i\lambda\kappa^{-(d+2)/2}
       \big(\langle f(\,\cdot/\kappa),\xi\rangle - \rho\kappa^d\int_{\R^d}f(x)\,dx
      \big)\big]\Big]
\]
\[
       = e^{-\lambda^2(\rho-\rho_c)\langle f, (-\beta\Delta)_{\Lambda}^{-1}f \rangle
       +o(1)} \ .
\]
Then setting  $ \lambda := t/\sqrt{2(\rho-\rho_c)\langle f,
(-\beta\Delta)_{\Lambda}^{-1}f \rangle} $,
we finally obtain the limit:
\begin{equation*}
\mathbb{E}_{\nu_{\rho}}\big[e^{itX_{\kappa}}\big] \to e^{-t^2/2} \ \ ,
\end{equation*}
which finishes the proof of the Central Limit Theorem. \hfill$\square$
\begin{remark}
The above calculations show that the value of the \textit{variation} that we need to
normalize the limiting random variable, is contributed from the measure
$\mu_{K^{\beta},(\rho - \rho_c)}$.
\end{remark}
%\bigskip

Before to pass to the \textit{Large Deviation Principle}, we prove the following lemma.

%%%%%%%%%%%%%%%%%%%%%%%%%%% LEMMA L-f %%%%%%%%%%%%%%%%%%%%%%%%%%%%%%
\begin{lem}
Let $\|\sqrt f (-\beta\Delta)_{\Lambda}^{-1}\sqrt f\| <1 $.
Then $ - \beta\Delta - f $ is a self-adjoint operator, which satisfies the property {\rm{:}}
Spec$\, (- \beta\Delta - f) \subset [0, \infty) $.
Moreover, the operator $\displaystyle (-\beta\Delta - f)^{-1}_{\Lambda}$ is bounded and
we have:
\begin{equation}
\langle \sqrt f, [ 1 -\sqrt f(-\beta\Delta)^{-1}_{\Lambda}
\sqrt f]^{-1}\sqrt f\rangle = \int_{\R^d}f(x)\, dx
+ \langle f, (-\beta\Delta -f)^{-1}_{\Lambda} f\rangle \ .
\label{l-f}
\end{equation}
\label{L-f}
\end{lem}
%%%%%%%%%%%%%%%%%%%%%%%%%%%%%%%%%%%%%%%%%%%%%%%%%%%%%%%%%%%%%%%%%%%%
{\sl Proof }: Since the operator $-\beta\Delta$ is self-adjoint, the spectrum
Spec$\, (- \beta\Delta) \subset [0, \infty) $ and $f$ is a bounded
function, it is obvious that $-\beta\Delta -f$ is self-adjoint
and $ (\delta -\beta\Delta)^{-1} $ is bounded non-negative operator
for arbitrary $ \delta > 0$. Since $ f \geqslant 0$ and $ \supp f \subset \Lambda $,
it is also obvious that
\[
     0 \leqslant \sqrt f  (\delta -\beta\Delta)^{-1}\sqrt f
         \leqslant  \sqrt f (-\beta\Delta)^{-1}_{\Lambda} \sqrt f \ .
\]
Together with the assumption $\|\sqrt f (-\beta\Delta)_{\Lambda}^{-1}\sqrt f\| <1$,
the operator
\begin{equation}
   S:= (\delta -\beta\Delta)^{-1} + (\delta -\beta\Delta)^{-1}\sqrt f
       \sum_{n=0}^{\infty}(\sqrt f  (\delta -\beta\Delta)^{-1}\sqrt f )^n
      \sqrt f  (\delta -\beta\Delta)^{-1}
\label{S}
\end{equation}
is a bounded non-negative operator. On the other hand, one can check that
\[
    (\delta -\beta\Delta -f )S = I  \quad \mbox{and} \quad
       S(\delta -\beta\Delta -f) = I_{\mbox{Dom}(\Delta)} \ ,
\]
which implies that $ S = (\delta -\beta\Delta -f)^{-1}$.
Thus, we have $ -\delta \not\in $ Spec$\, (- \beta\Delta -f)$, i.e.,
Spec$\, - \beta\Delta -f \subset [0, \infty) $.
Let $\{E(\lambda)\}$ be the spectral decomposition of the operator $ (-\beta\Delta -f) $.
Then $E(-0) = 0$. Moreover, $E(0) = 0$ holds.
Indeed, if one supposes the contrary, then there exists a $\psi \ne 0$ such that
\[
     \psi \in E(0)L^2(\R^d) \quad \mbox{ and } \quad (-\beta\Delta -f )\psi = 0 \ .
\]
Thus, we have $f\psi = -\beta\Delta\psi$, which implies that
\[
        f\psi \in \mbox{Ran}\,(-\beta\Delta) = \mbox{Dom}\,(-\beta\Delta)^{-1}
\]
and
\[
        \psi = (-\beta\Delta)^{-1}f\psi.
\]
Hence we get $ \sqrt f\psi = \sqrt f(-\beta\Delta)^{-1} f\psi =
(\sqrt f(-\beta\Delta)_{\Lambda}^{-1}\sqrt f)\sqrt f\psi$.
This contradicts the estimate $\|\sqrt f(-\beta\Delta)_{\Lambda}^{-1}\sqrt f\| <1 $
because $ \sqrt f\psi \in L^2(\R^d)$ belong to the eigenvalue 1 of the operator
$ \sqrt f (-\beta\Delta)_{\Lambda}^{-1}\sqrt f $.
Therefore, we obtain densely defined non-negative self-adjoint operator:
\[
         (-\beta\Delta -f)^{-1} := \int_0^{\infty}\frac{dE(\lambda)}{\lambda}.
\]
%\medskip
The boundedness of $(-\beta\Delta-f)_{\Lambda}^{-1}$ follows from the estimates:
\begin{eqnarray*}
     \|(-\beta\Delta-f)_{\Lambda}^{-1}\| &=& \sup_{\|\phi \|_2 =1}
     \int_0^{\infty}\frac{d \langle\chi_{\Lambda}\phi,
        E(\lambda)\chi_{\Lambda}\phi\rangle}{\lambda}
\\
     &=&   \sup_{\|\phi \|_2 =1}\lim_{\delta\downarrow 0}
     \int_0^{\infty}\frac{d \langle\chi_{\Lambda}\phi,
        E(\lambda)\chi_{\Lambda}\phi\rangle}{\delta + \lambda}
     = \sup_{\|\phi \|_2 =1}\lim_{\delta\downarrow 0} \langle \chi_{\Lambda}\phi,
           S \chi_{\Lambda}\phi\rangle
\\
      &\leqslant& \sup_{\|\phi \|_2 =1}
           \Big[ \langle \phi, (-\beta\Delta)_{\Lambda}^{-1} \phi\rangle
            + \frac{\|\sqrt f(\delta-\beta\Delta)^{-1}\chi_{\Lambda}\phi\|_2^2}
                 {1- \| \sqrt f(-\beta\Delta)_{\Lambda}^{-1}\sqrt f \|}\Big]
\\
      &\leqslant& \|(-\beta\Delta)_{\Lambda}^{-1}\| +
    \frac{\| f\|_{\infty}\|(\delta-\beta\Delta)^{-1}_{\Lambda}\|^2}
                 {1- \| \sqrt f(-\beta\Delta)_{\Lambda}^{-1}\sqrt f \|}
         \; < \; \infty.
\end{eqnarray*}

To derive equation (\ref{l-f}), we exploit the operator (\ref{S}) for $\delta \downarrow 0$:
\[
    \langle f,  (\delta-\beta\Delta -f)^{-1}f\rangle =
      \langle \sqrt f, \sum_{n=1}^{\infty}(\sqrt f (\delta-\beta\Delta)^{-1}
      \sqrt f)^n \sqrt f\rangle
\]
\[
   =   - \langle \sqrt f, \sqrt f\rangle +
        \langle \sqrt f, ( 1 - \sqrt f (\delta-\beta\Delta)^{-1}\sqrt f)^{-1} \sqrt f\rangle
\]
\[
    \longrightarrow -\int f\,dx +  \langle \sqrt f,
     ( 1 - \sqrt f (-\beta\Delta)_{\Lambda}^{-1}\sqrt f)^{-1} \sqrt f\rangle \ ,
\]
where we used the convergence
\[
         \sqrt f (\delta-\beta\Delta)^{-1}\sqrt f \to
          \sqrt f (-\beta\Delta)^{-1}_{\Lambda}\sqrt f
\]
in the \textit{operator norm}. The latter is a direct consequence of the spectral theorem and the
dominated convergence theorem. On the other hand, we notice that for $\delta \downarrow 0$ one gets
by the monotone convergence theorem the limit:
\[
   \langle f,  (\delta-\beta\Delta -f)^{-1}f\rangle  =
    \int_0^{\infty}\frac{d \langle\chi_{\Lambda} f,
        E(\lambda)\chi_{\Lambda} f \rangle}{\delta + \lambda}
\]
\[
    \longrightarrow \int_0^{\infty}\frac{d \langle\chi_{\Lambda} f,
        E(\lambda)\chi_{\Lambda} f \rangle}{ \lambda}
      = \langle f,  (-\beta\Delta -f)^{-1}_{\Lambda}f\rangle \ \ .
\]
Therefore,  the equality (\ref{l-f}) is proven. \hfill$\square$

%\medskip

%%%%%%%%%%%%%%%%%%%%%%%%% THM LDP1 %%%%%%%%%%%%%%%%%%%%%%%%%%%%%
\begin{thm}
For any bounded measurable function $f \geqslant 0$ with bounded support
and for any bounded measurable subset $\Lambda $ of $\R^d$ satisfying $ \supp f \subset \Lambda$
we have the following limits:
\[
       P(t) := \lim_{\kappa \to \infty}\frac{1}{\kappa^{d-2}}
          \log \int_{Q(\R^d)}e^{t\kappa^{-2}\langle f(\,\cdot/\kappa), \xi\rangle}
           \nu_{\rho}(d\xi)
\]
\[
        = \begin{cases}
               \rho t \int_{\R^d}f(x)\, dx + (\rho -\rho_c)t^2
                 \langle f,( -\beta\Delta -tf)_{\Lambda}^{-1}f\rangle & {\rm{ for }} \;
                   t \in (-\infty, \|\sqrt f(-\beta \Delta)_{\Lambda}^{-1}\sqrt f\|^{-1}),
           \\
                \infty & {\rm{ for }} \;
                   t \in [\|\sqrt f(-\beta \Delta)_{\Lambda}^{-1}\sqrt f\|^{-1}, \infty)\ .
         \end{cases}
\]
\label{LDP1}
\end{thm}
%%%%%%%%%%%%%%%%%%%%%%%%%%%%%%%%%%%%%%%%%%%%%%%%%%%%%%%%%%%%%%%%%%%%%%%%%%%%%%%%%%%%%%%%%%%%%%%%%%%%%
\begin{remark}
{\rm{(1)}} If $ \displaystyle   t < \|\sqrt f(\beta \Delta)_{\Lambda}^{-1}\sqrt f\|^{-1}$, then 
from Lemma \ref{L-f} we obtain the expression for the function $P(t)$ :
\begin{equation}\label{P}
P(t) = \rho_c t \int_{\R^d}f(x)\, dx + (\rho -\rho_c)t
\langle \sqrt f,[ 1 -t\sqrt f(-\beta\Delta)^{-1}_{\Lambda}
\sqrt f]^{-1}\sqrt f\rangle \ .
\end{equation}

{\rm{(2)}} By Lemmas \ref{KHS} and \ref{K2Lf} the operator $ \sqrt f(-\beta\Delta)^{-1}_{\Lambda}\sqrt f $ is 
a non-negative compact operator. Let $ \{\varphi_n\} $ be a CONS of $L^2(\R^2)$, which consists of the eigenfunctions
of this operator. We order the corresponding eigenvalues as
\[
        \lambda_1 \geqslant \lambda_2 \geqslant \cdots \geqslant \lambda_n \geqslant \cdots
         \geqslant 0
\]
Then the Perron-Frobenius theorem yields
\[
         \lambda_1 = \|\sqrt f(-\beta\Delta)^{-1}_{\Lambda}\sqrt f \|
    \quad \mbox{ and } \quad  \langle \sqrt f, \varphi_1\rangle > 0 .
\]
By the remark {\rm{(1)}}, we obtain:
\[
       P(t) = \rho_c \ t \int_{\R^d}f(x)\, dx + (\rho -\rho_c)\ t \ \sum_{n=1}^{\infty}
                 \frac{|\langle \sqrt f, \varphi_n\rangle|^2}{1-t\lambda_n}
    \quad {\rm{for}} \quad t < \lambda_1^{-1} \ ,
\]
which ensures the (essential) smoothness of $P$:
\[
       P \, \mbox{ is a } \, C^{\infty} \mbox{ function on } (-\infty, \lambda_1^{-1})
 \quad  {\rm{ and }} \quad   \lim_{t\uparrow \lambda_1^{-1}}P(t) = \infty \ .
\]

{\rm{(3)}} Below we prove that the limits of the function $P$ for the components $\mu_{K,\rho}$ and 
$\mu_K^{(det)}$ of the boson random point processes have the following forms:
\[
       P_{K^{\beta}, \, \rho}(t) := \lim_{\kappa \to \infty}\frac{1}{\kappa^{d-2}}
          \log \int_{Q(\R^d)} e^{t\kappa^{-2}\langle f(\,\cdot/\kappa), \xi\rangle}
           \mu_{K^{\beta}, \, \rho}(d\xi)
\]
\[
        = \begin{cases}
               \rho t \langle \sqrt f,[ 1 -t\sqrt f(-\beta\Delta)^{-1}_{\Lambda}
         \sqrt f]^{-1}\sqrt f\rangle & {\rm{ for }} \;
                   t \in (-\infty, \|\sqrt f(-\beta \Delta)_{\Lambda}^{-1}\sqrt f\|^{-1}) \ ,
           \\
                \infty & {\rm{ for }} \;
                   t \in [\|\sqrt f(-\beta \Delta)_{\Lambda}^{-1}\sqrt f\|^{-1}, \infty) \ ,
         \end{cases}
\]
and
\[
       P_{K^{\beta}}^{(det)}(t) := \lim_{\kappa \to \infty}\frac{1}{\kappa^{d-2}}
          \log \int_{Q(\R^d)}e^{t\kappa^{-2}\langle f(\,\cdot/\kappa), \xi\rangle}
           \mu_{K^{\beta}}^{(det)}(d\xi)
\]
\[
        = \begin{cases}
               \rho_c \, t \, \int_{\R^d}f(x)\, dx & {\rm{ for }} \;
                   t \in (-\infty, \|\sqrt f(-\beta \Delta)_{\Lambda}^{-1}\sqrt f\|^{-1}) \ ,
           \\
                \infty & {\rm{ for }} \;
                   t \in [\|\sqrt f(-\beta \Delta)_{\Lambda}^{-1}\sqrt f\|^{-1}, \infty) \ .
         \end{cases} 
\]
\label{remLDP}
\end{remark}
{\sl Proof }(of Theorem \ref{LDP1}):
The proof consists of two parts corresponding to  $t<0$ and  $t>0$. ( The case $t=0$ is obvious.)

$ 1^{\circ}$. For $t<0$, it is enough to show that
\[
    P(-1) =  \lim_{\kappa \to \infty}\frac{1}{\kappa^{d-2}}
          \log \int_{Q(\R^d)}e^{-\kappa^{-2}\langle f(\,\cdot/\kappa), \xi\rangle}
           \nu_{\rho}(d\xi)
\]
\[
        =  - \rho  \int_{\R^d}f(x)\, dx + (\rho -\rho_c)
                 \langle f,( -\beta\Delta +f)_{\Lambda}^{-1}f\rangle.
\]
To this end  notice that from (\ref{chdet}) and (\ref{chrho}), together with the unitary transformation $U_{\kappa}$, 
one obtains the representation:
\[
        \frac{1}{\kappa^{d-2}}
          \log \int_{Q(\R^d)}e^{-\kappa^{-2}\langle f(\,\cdot/\kappa), \xi\rangle}
           \nu_{\rho}(d\xi)
\]
\[
         = -\frac{\rho-\rho_c}{\kappa^{d-2}}\Big\langle
       U_{\kappa} \sqrt{1-e^{-\kappa^{-2}f(\,\cdot/\kappa)}}, U_{\kappa}
       (1+\sqrt{1-e^{-\kappa^{-2}f(\,\cdot/\kappa)}}K^{\beta}_{\kappa\Lambda}
       \sqrt{1-e^{-\kappa^{-2}f(\,\cdot/\kappa)}})^{-1}
      \sqrt{1-e^{-\kappa^{-2}f(\,\cdot/\kappa)}}\Big\rangle
\]
\[
         - \frac{1}{\kappa^{d-2}} \log \Det\big[1+U_{\kappa}\big(1-e^{-\kappa^{-2}
        f(\,\cdot/\kappa)} \big) K^{\beta}_{\kappa\Lambda}U_{\kappa}^{-1}\big]
\]
\[
      = -(\rho-\rho_c)\big\langle
        \sqrt{f^{(-)}_{\kappa}}, \big(1+\sqrt{f^{(-)}_{\kappa}}
          \kappa^{-2}K^{\beta/\kappa^{2}}_{\Lambda} \sqrt{f^{(-)}_{\kappa}}\big)^{-1}
              \sqrt{f^{(-)}_{\kappa}}\big\rangle
\]
\[
        -\frac{1}{\kappa^d}\Tr \big[ \sqrt{f^{(-)}_{\kappa}}
        K^{\beta/\kappa^{2}}_{\Lambda} \sqrt{f^{(-)}_{\kappa}} \big]
         - \frac{1}{\kappa^{d-2}}\log\Det_2\big[1+f^{(-)}_{\kappa}\kappa^{-2}
              K^{\beta/\kappa^{2}}_{\Lambda}\big] \ ,
\]
see Definition \ref{defPM}. Then we apply Lemmas \ref{fmp}, \ref{K2Lf} to the first term, Lemma \ref{KTR} to 
the second term and Lemma \ref{KHS} with (\ref{Det_2}) to the third term to obtain:
\[
     P(-1) =    - (\rho -\rho_c) \langle \sqrt f, [ 1 +\sqrt f(-\beta\Delta)^{-1}_{\Lambda}
         \sqrt f]^{-1}\sqrt f\rangle -\rho_c  \int_{\R^d}f(x)\, dx \ .
\]

Now it is sufficient to check the identity:
\begin{equation}
     \langle \sqrt f, [ 1 +\sqrt f(-\beta\Delta)^{-1}_{\Lambda}
         \sqrt f]^{-1}\sqrt f\rangle = \int_{\R^d}f(x)\, dx -
      \langle  f, ( -\beta\Delta +f )^{-1}_{\Lambda} f\rangle \ .
\label{f1}
\end{equation}
Note that the inequality
\[
        (-\beta\Delta +f)_{\Lambda}^{-1} \leqslant (-\beta\Delta)_{\Lambda}^{-1}
\]
yields that  $ \displaystyle (-\beta\Delta +f)_{\Lambda}^{-1} $ is bounded.
Since the operators $ (\epsilon -\beta\Delta  )^{-1}, (\epsilon -\beta\Delta +f )^{-1}$
are bounded and non-negative for any $\epsilon >0$, we get
\[
    \sqrt f (\epsilon -\beta\Delta  )^{-1}\sqrt f -
          \sqrt f(\epsilon -\beta\Delta +f )^{-1}\sqrt f
     = \sqrt f(\epsilon -\beta\Delta  )^{-1} \sqrt f \sqrt f
      (\epsilon -\beta\Delta +f )^{-1}\sqrt f \ .
\]
It gives
\[
     \sqrt f (\epsilon -\beta\Delta  )^{-1}\sqrt f  =
   (1+   \sqrt f (\epsilon -\beta\Delta  )^{-1}\sqrt f)
               \sqrt f(\epsilon -\beta\Delta +f )^{-1}\sqrt f \ ,
\]
which implies
\[
   1- (1+\sqrt f(\epsilon -\beta\Delta )^{-1}\sqrt f)^{-1}
       =  \sqrt f(\epsilon -\beta\Delta +f )^{-1}\sqrt f \ .
\]
Hence, to verify (\ref{f1}), it is enough to prove that
\begin{eqnarray}
    \sqrt f(\epsilon -\beta\Delta +f )^{-1}\sqrt f &\to
    \sqrt f( -\beta\Delta +f )^{-1}_{\Lambda}\sqrt f    \qquad \mbox{weakly} \ ,
\label{f2} \\
     \sqrt f(\epsilon -\beta\Delta )^{-1}\sqrt f &\to
    \sqrt f( -\beta\Delta )^{-1}_{\Lambda}\sqrt f    \qquad \mbox{ in norm} \ .
\label{f3}
\end{eqnarray}

To show (\ref{f2}), let $\{ E(\lambda)\} $ be the spectral decomposition
of $ -\beta\Delta +f $.
Since
\[
   \int_0^{\infty}\frac{d \langle \sqrt f \phi, E(\lambda)\sqrt f \phi\rangle}{\lambda}
  = \langle \phi, \sqrt f( -\beta\Delta +f)^{-1}_{\Lambda}\sqrt f \phi\rangle
\]
\[
      \leqslant \langle \phi, \sqrt f( -\beta\Delta)^{-1}_{\Lambda}\sqrt f \phi\rangle
          < \infty
\]
holds for $\phi \in L^2(\R^d)$, the dominated convergence theorem yields the limit:
\[
       | \langle \phi, \sqrt f( -\beta\Delta +f)^{-1}_{\Lambda}\sqrt f \phi\rangle
    - \langle \phi, \sqrt f(\epsilon -\beta\Delta +f)^{-1}_{\Lambda}\sqrt f \phi\rangle |
\]
\[
     = \int_0^{\infty}\Big(\frac{1}{\lambda} - \frac{1}{\lambda + \epsilon} \Big)
                d \langle \sqrt f \phi, E(\lambda)\sqrt f \phi\rangle \to 0.
\]

To show (\ref{f3}), we use the Fourier transformation. Put
\[
      \| \sqrt f( -\beta\Delta)^{-1}_{\Lambda}\sqrt f -
        \sqrt f( \epsilon -\beta\Delta )^{-1}_{\Lambda}\sqrt f \|
\]
\[
       = \sup_{\|\phi\|_2 =1}\int_{\R^d}\frac{\epsilon |\widetilde{\sqrt f \phi}(p)|^2}
                {\beta|p|^2(\epsilon +\beta |p|^2)} \, dp =: D.
\]
When $ d>4$, we obtain that
\begin{eqnarray*}
     D &\leqslant& \sup_{\|\phi\|_2 =1}\int_{|p|<1}
         \frac{\epsilon \|\widetilde{\sqrt f \phi}\|_{\infty}^2}{\beta^2|p|^4} dp
      + \sup_{\|\phi\|_2 =1}\int_{|p|\geqslant 1}\frac{\epsilon |\widetilde{\sqrt f \phi}(p)|^2}
                {\beta^2}dp
\\
    &\leqslant& \frac{\epsilon}{\beta^2}(c_d \|f\|_1 + \|f\|_{\infty}) \to 0 \ ,
\end{eqnarray*}
for $ \epsilon \to 0$. When $ 2<d<4$, we get
\[
      D \leqslant \sup_{\|\phi\|_2 =1}\int_{\R^d}
    \frac{\epsilon \|\widetilde{\sqrt f \phi}\|_{\infty}^2}{\beta|p|^2(\epsilon +\beta|p|^2)}
     \, dp
\]
\[
     \leqslant \frac{\epsilon^{(d-2)/2}}{\beta^{d/2}}\int_{\R^d}\frac{\|f\|_1 \,d\tilde p}
      {(2\pi)^d|\tilde p|^2(1+ |\tilde p|^2)}  \to 0,
\]
as $\epsilon \to 0$. Here we used the bounds
$ \|\widetilde{\sqrt f \phi}\|_{\infty} \leqslant (2\pi)^{-d/2} \|f\|_1^{1/2}\|\phi\|_2$
and
$ \|\widetilde{\sqrt f \phi}\|_2 \leqslant \|f\|_{\infty}^{1/2}\|\phi\|_2 $,
and changed the integral variable $ p = \sqrt{\epsilon/\beta}\tilde p$ in the latter integral.

Similarly for $ d=4$, we obtain the limit:
\begin{eqnarray*}
     D &\leqslant& \sup_{\|\phi\|_2 =1}\int_{|p|<1}
         \frac{\epsilon \|\widetilde{\sqrt f \phi}\|_{\infty}^2}
        {\beta|p|^2(\epsilon +\beta|p|^2)} dp
      + \sup_{\|\phi\|_2 =1}\int_{|p|\geqslant 1}\frac{\epsilon |\widetilde{\sqrt f \phi}(p)|^2}
                {\beta^2}dp
\\
     &\leqslant& \frac{\epsilon^{(d-2)/2}}{\beta^{d/2}}\int_{|\tilde p|<\sqrt{\beta/\epsilon}}
     \frac{\|f\|_1 \,d\tilde p}{(2\pi)^d|\tilde p|^2(1+ |\tilde p|^2)}
     + \frac{\epsilon}{\beta^2}\|f\|_{\infty}
\\
    &\leqslant& c \|f\|_1 \frac{\epsilon}{\beta^{2}}
        \log\Big(1+\frac{\beta}{\epsilon}\Big) + \frac{\epsilon}{\beta^2}\|f\|_{\infty}
     \to 0
\end{eqnarray*}
when  $\epsilon \to 0$.

\bigskip

$ 2^{\circ}$. For $t>0$, It is enough to show
\[
       P(1) = \begin{cases}
               \rho  \int_{\R^d}f(x)\, dx + (\rho -\rho_c)
                 \langle f,( -\beta\Delta -f)_{\Lambda}^{-1}f\rangle & \mbox{ for } \;
                    \|\sqrt f(-\beta \Delta)_{\Lambda}^{-1}\sqrt f\| <1 \ ,
           \\
                \infty & \mbox{ for } \;
                   \|\sqrt f(-\beta \Delta)_{\Lambda}^{-1}\sqrt f\| \geqslant 1 \ .
         \end{cases}
\]

When $ \|\sqrt f(-\beta \Delta)_{\Lambda}^{-1}\sqrt f\| <1 $, then by Lemma \ref{K2Lf} and Lemma \ref{U}
we have
\[
       \|\sqrt{ f^{(+)}_{\kappa}}\kappa^{-2}K^{\beta/\kappa^{2}}_{\Lambda}
     \sqrt{ f^{(+)}_{\kappa}}\|  =
 \|\sqrt{e^{\kappa^{-2}f(\cdot/\kappa)}-1}\ K^{\beta}_{\kappa\Lambda} \ 
 \sqrt{e^{\kappa^{-2}f(\cdot/\kappa)}-1}\| < 1
\]
for $\kappa$ large enough, see Definition \ref{defPM}.
We also use Lemma \ref{U} and Theorem \ref{thmexp} (2),(4) to obtain the representation:
\[
    \frac{1}{\kappa^{d-2}}
          \log \int_{Q(\R^d)}e^{\kappa^{-2}\langle f(\,\cdot/\kappa), \xi\rangle}
           \nu_{\rho}(d\xi)
\]
\[
    = \frac{\rho-\rho_c}{\kappa^{d-2}}\Big\langle
       U_{\kappa} \sqrt{e^{\kappa^{-2}f(\,\cdot/\kappa)}-1}, U_{\kappa}
       (\sqrt{e^{\kappa^{-2}f(\,\cdot/\kappa)}-1}K^{\beta}_{\kappa\Lambda}
       \sqrt{e^{\kappa^{-2}f(\,\cdot/\kappa)}-1})^{-1}
      \sqrt{e^{\kappa^{-2}f(\,\cdot/\kappa)}-1}\Big\rangle
\]
\[
         - \frac{1}{\kappa^{d-2}} \log \Det\big[1-U_{\kappa}
         \sqrt{e^{\kappa^{-2}f(\,\cdot/\kappa)}-1}K^{\beta}_{\kappa\Lambda}
       \sqrt{e^{\kappa^{-2}f(\,\cdot/\kappa)}-1}U_{\kappa}^{-1}\big]
\]
\[
      = (\rho-\rho_c)\big\langle
        \sqrt{f^{(+)}_{\kappa}}, \big(1-\sqrt{f^{(+)}_{\kappa}}
          \kappa^{-2}K^{\beta/\kappa^{2}}_{\Lambda} \sqrt{f^{(+)}_{\kappa}}\big)^{-1}
          \sqrt{f^{(+)}_{\kappa}}\big\rangle
\]
\[
        +\frac{1}{\kappa^d}\Tr[ f^{(+)}_{\kappa}K^{\beta/\kappa^{2}}_{\Lambda}]
         - \frac{1}{\kappa^{d-2}}\log\Det_2\big[1-f^{(+)}_{\kappa}\kappa^{-2}
              K^{\beta/\kappa^{2}}_{\Lambda}\big] \ .
\]
Applying Lemma \ref{K2Lf}, \ref{fmp} to the first term, Lemma \ref{KTR}, \ref{fmp}
to the second term and Lemma \ref{KHS} to the third term, we get
\[
   P(1) = (\rho -\rho_c) \langle \sqrt f, [ 1 -\sqrt f(-\beta\Delta)^{-1}_{\Lambda}
         \sqrt f]^{-1}\sqrt f\rangle +  \rho_c  \int_{\R^d}f(x)\, dx .
\]
Then the  Lemma \ref{L-f} proves the case $ \|\sqrt f(-\beta \Delta)_{\Lambda}^{-1}\sqrt f\| <1 $.

When $\|\sqrt f(-\beta\Delta)_{\Lambda}^{-1}\sqrt f\| >1$,
we apply $U_{\kappa}$ and Lemmas \ref{K2Lf}, \ref{U} to find that
\[
      \| \sqrt{e^{f(\,\cdot/\kappa)/\kappa^2}-1} K^{\beta}_{\kappa\Lambda}
       \sqrt{e^{f(\,\cdot/\kappa)/\kappa^2}-1}\| \geqslant 1 \ ,
\]
for $\kappa$ large enough. Therefore, we get from Theorem \ref{thmexp}(2),(4) that
\[
    \lim_{\kappa \to \infty} \int_{Q(\R^d)}e^{\langle f(\,\cdot/\kappa)/\kappa^2 ,\xi\rangle}
     \nu_{\rho}(d\xi) = \infty \ .
\]

When $\|\sqrt f(-\beta\Delta)_{\Lambda}^{-1}\sqrt f\| = 1$, then applying 
Lemma \ref{K2Lf} and transformation $U_{\kappa}$, we find for large $\kappa$ the estimate:
\[
       \| \sqrt{e^{f(\,\cdot/\kappa)/\kappa^2}-1} K^{\beta}_{\kappa\Lambda}
       \sqrt{e^{f(\,\cdot/\kappa)/\kappa^2}-1}\|
      = \| \sqrt{f^{(+)}_{\kappa}}\kappa^{-2}K^{\beta/\kappa^2}_{\Lambda}
        \sqrt{f^{(+)}_{\kappa}}\| \geqslant 1 - c\kappa^{-2} \ .
\]
In fact, it is enough to consider the case where the above quantity is smaller than 1.
In this case Lemmas \ref{fmp}, \ref{K2Lf} and \ref{Linv} yield
\begin{eqnarray*}
  & &|\langle \sqrt{f^{(+)}_{\kappa}}, (\sqrt{f^{(+)}_{\kappa}}\kappa^{-2}
       K^{\beta/\kappa^2}_{\Lambda}\sqrt{f^{(+)}_{\kappa}})^n
       \sqrt{f^{(+)}_{\kappa}} \rangle -
       \langle \sqrt{f}, (\sqrt{f}(-\beta\Delta)_{\Lambda}^{-1}\sqrt{f})^n
       \sqrt{f}\rangle |
\\
  & \leqslant & |\langle \sqrt{f^{(+)}_{\kappa}} -\sqrt f, (\sqrt{f^{(+)}_{\kappa}}\kappa^{-2}
       K^{\beta/\kappa^2}_{\Lambda}\sqrt{f^{(+)}_{\kappa}})^n
       \sqrt{f^{(+)}_{\kappa}} \rangle|
\\
  & & + |\langle \sqrt{f} , (\sqrt{f^{(+)}_{\kappa}}\kappa^{-2}
       K^{\beta/\kappa^2}_{\Lambda}\sqrt{f^{(+)}_{\kappa}})^n
       (\sqrt{f^{(+)}_{\kappa}}-\sqrt f) \rangle|
\\
  & & + |\langle \sqrt{f}, \{(\sqrt{f^{(+)}_{\kappa}}\kappa^{-2}
       K^{\beta/\kappa^2}_{\Lambda}\sqrt{f^{(+)}_{\kappa}})^n
        - (\sqrt{f}(-\beta\Delta)_{\Lambda}^{-1}\sqrt{f})^n\}
       \sqrt{f}\rangle |
\\
  &\leqslant & \frac{\|f\|_{\infty}}{\kappa^2}(1+e^{\|f\|_{\infty}/\kappa^2})
      e^{\|f\|_{\infty}/\kappa^2}\langle \sqrt{f}, \sqrt{f}\rangle
\\
  & & + n\langle \sqrt{f}, \sqrt{f}\rangle
        \| \sqrt{f^{(+)}_{\kappa}} \kappa^{-2}
       K^{\beta/\kappa^2}_{\Lambda}\sqrt{f^{(+)}_{\kappa}}
        - \sqrt{f}(-\beta\Delta)_{\Lambda}^{-1}\sqrt{f} \|
\\
 &\leqslant & c \frac{n+1}{\kappa^2}.
\end{eqnarray*}
This estimate together with Theorem \ref{thmexp}(2), give the limit:
\begin{eqnarray*}
    &&\frac{1}{\kappa^{d-2}}\log\int_{Q(\R^d)}e^{\langle f(\,\cdot/\kappa)/\kappa^2,
     \xi\rangle} \mu_{K^{\beta},(\rho-\rho_c)}(d\xi)
\\
     & = & \frac{\rho-\rho_c}{\kappa^{d-2}} \big\langle
      \sqrt{e^{f(\,\cdot/\kappa)/\kappa^2}-1},
     \big(1- \sqrt{e^{f(\,\cdot/\kappa)/\kappa^2}-1} K^{\beta}_{\kappa\Lambda}
       \sqrt{e^{f(\,\cdot/\kappa)/\kappa^2}-1}\big)^{-1}
       \sqrt{e^{f(\,\cdot/\kappa)/\kappa^2}-1}\big\rangle
\\
    &=& (\rho-\rho_c) \sum_{n=0}^{\infty} \langle \sqrt{f^{(+)}_{\kappa}},
      (\sqrt{f^{(+)}_{\kappa}}\kappa^{-2}
       K^{\beta/\kappa^2}_{\Lambda}\sqrt{f^{(+)}_{\kappa}})^n
       \sqrt{f^{(+)}_{\kappa}} \rangle
\\
   & \geqslant& (\rho-\rho_c) \sum_{n=0}^{\infty} \big\{
      \langle \sqrt{f}, (\sqrt{f}(-\beta\Delta)_{\Lambda}^{-1}\sqrt{f})^n
       \sqrt{f}\rangle
\\
   && - |\langle \sqrt{f^{(+)}_{\kappa}},
      (\sqrt{f^{(+)}_{\kappa}}\kappa^{-2}
       K^{\beta/\kappa^2}_{\Lambda}\sqrt{f^{(+)}_{\kappa}})^n
       \sqrt{f^{(+)}_{\kappa}} \rangle
      -  \langle \sqrt{f}, (\sqrt{f}(-\beta\Delta)_{\Lambda}^{-1}\sqrt{f})^n
       \sqrt{f}\rangle|
           \big\}\vee 0
\\
  & \geqslant& (\rho-\rho_c) \sum_{n=0}^{\infty}\big\{ |\langle \varphi,\sqrt f\rangle|^2
   -  c\frac{n+1}{\kappa^2} \big\}\vee 0
    \geqslant (\rho-\rho_c)\frac{|\langle \varphi,\sqrt f\rangle|^4\kappa^2}{2c} \to \infty \ .
\end{eqnarray*}
when  $\kappa \to \infty$. Here we applied $U_{\kappa}$ in the second equality, and then the fact that $\varphi $ 
is the eigenfunction of the operator $\sqrt{f}(-\beta\Delta)_{\Lambda}^{-1}\sqrt{f}$ with the largest eigenvalue 1.
Note that $\langle \sqrt f, \varphi \rangle >0 $.
In fact, since the integral kernel of this operator is positive on the set $\{f>0\}$, one gets:
$\varphi > 0 \;a.e. \mbox{ on } \{f>0\}$, c.f. Remark \ref{remLDP}(2).

The corresponding estimate for $\mu_{K^{\beta}}^{(det)}$ is straightforward. \hfill $\square$

\smallskip

\noindent Recall that the Fenchel-Legendre transformation of the function $P$ has the form:
\[
        I(s) := \sup_{s \in \R} \big(st -P(t) \big)
\]
By virtue of the G\"artner-Ellis theorem (see e.g. \cite{DZ}), we obtained the following
large deviation principle.

%%%%%%%%%%%%%%%%%%%%%%%%%%%% Theorem {LDP2} %%%%%%%%%%%%%%%%%%%%%
\begin{thm}[Large Deviation Principle]
The random variable $\displaystyle \langle f(\,\cdot/\kappa)/\kappa^2, \xi\rangle $ satisfies
in the condensation regime $\rho > \rho_c$ the large deviation principle with a bona fide rate function $I$:
\[
      \limsup_{\kappa \to \infty}\frac{1}{\kappa^{d-2}}\log \nu_{\rho}
       \Big[ \Big\langle \frac{1}{\kappa^d}f\Big(\frac{.}{\kappa}\Big), \xi \Big\rangle
       \in F \Big] \leqslant -\inf_{s\in F}I(s)
        \qquad {\rm{ for arbitrary closed }} \; F \subset \R
\]
and
\[
       \liminf_{\kappa \to \infty}\frac{1}{\kappa^{d-2}}\log \nu_{\rho}
       \Big[ \Big\langle \frac{1}{\kappa^d}f\Big(\frac{.}{\kappa}\Big), \xi \Big\rangle
       \in G \Big] \geqslant -\inf_{s\in G}I(s)
        \qquad {\rm{ for arbitrary open }} \; G \subset \R \ .
\]
\label{LDP2}
\end{thm}
\begin{remark}\label{shift}
Note that contribution of the point processus $\mu_{K^\beta}^{(det)}$ to the large deviation
property is in a sense marginal, since it only shifts the variable $s$ of the rate function
$I$ see (\ref{P}). Taking into account the central limit theorem, we see that the characteristic
feature of the limit theorems for the ideal boson gas in the presence of the Bose-Einstein condensation  
is reflected by the convolution with a nontrivial component $\mu_{K^{\beta}, \rho}$. This gives 
$\nu_{\rho} = \mu_{K^{\beta}}^{(det)}\ast \mu_{K^{\beta},(\rho-\rho_c)}$.
\end{remark}
%%%%%%%%%%%%%%%%%%%%%%%%%% Abstract theory %%%%%%%%%%%%%%%%%%%%%%%%%%%%

\section{Conclusion}
%%%%%%%%%%%%%%%%%%%%%%%%%%%%%%%%%%%%%%%%%%%%%%%%%%%%%%%%%%%%%%%%%%%%%%%
To compare our results for the case: $\rho > \rho_c$ (BEC), we would like to mention here the corresponding 
results for the case $\rho < \rho_c$ (normal phase without condensation).

Let us put $K_z: =z G^{\beta}(1-z G^{\beta})^{-1}$ with $z \in (0,1) $, which
satisfies $ \rho = K_z(x,x)$ and $ \nu_{\rho} = \mu^{(det)}_{K_z} $.
Then for $\rho < \rho_c$ our theorems take the following form, see \cite{LLS, GLM, ST03}:
\begin{thm}[The law of large number]  For $\kappa \to \infty$ one has
\[
      \frac{1}{\kappa^d}\langle  f(\cdot/\kappa), \xi \rangle \longrightarrow
      \rho \int_{\R^d}f(x)\, dx \qquad  {\rm{in}} \quad L^2( Q(\R^d), \nu_{\rho}) \ .
\]
\label{5.1}
\end{thm}

\begin{thm}[The central limit theorem] For the random variables 
\[
      Z_{\kappa} = \frac{\langle  f(\cdot/\kappa), \xi \rangle -
                  \kappa^d \rho \int_{\R^d}f(x)\, dx}
          {\sqrt{K_z(x,x)+K_z^2(x,x)} \|f \|_2 \kappa^{d/2}},
\]
one gets the limit:
\[
     \lim_{\kappa \to \infty}\int_{Q(\R^d)}e^{itZ_{\kappa}} \nu_{\rho}(d\xi)
           =  e^{-t^2/2} \ .
\]
\label{5.2}
\end{thm}

\begin{thm}[Large deviation principle]
There exists a certain bona fide rate convex function $I':\R \mapsto [0, + \infty]$, such that
\[
     \limsup_{\kappa\to\infty}\frac{1}{\kappa^{d}} \log \nu_{\rho}
     \Big( \frac{1}{\kappa^d}\big\langle f\big(\cdot/\kappa\big),
    \xi\big\rangle \in F\Big) \leqslant -\inf_{s\in F}I'(s)
     \qquad \hphantom{L}
\quad {\rm{for any closed }} F \subset \R
\]
and
\[
     \liminf_{\kappa\to\infty}\frac{1}{\kappa^{d}} \log \nu_{\rho}
     \Big( \frac{1}{\kappa^d}\big\langle f\big(\cdot/\kappa\big),
    \xi\big\rangle \in G\Big) \geqslant -\inf_{s\in G}I'(s)
    \qquad \hphantom{L}
\quad {\rm{for any open }} G \subset \R 
\]
hold.
\label{5.3}
\end{thm}

We may summarize the difference between Theorems \ref{5.1}-\ref{5.3} and Theorems \ref{LLN}-\ref{LDP}
as follows. Let
\[
         D_{\kappa} = \frac{1}{\kappa^d}\langle  f(\cdot/\kappa), \xi \rangle,
\]
be the a random variable corresponding to empirical ``density" of particles localised in the region 
of length scale $\kappa$.

\smallskip

\noindent For the BEC case one gets:

\noindent (i) The random variable $D_{\kappa} $ converges for $\kappa \to \infty$ to its expectation value
$ m = \rho \int_{{\mathbb{R}}^d} f(x)\, dx$ in mean.

\noindent (ii) The law of the random variable $\kappa^{(d-2)/2}(D_{\kappa} -m)$ converges to normal 
distribution as $\kappa \to \infty$.

\noindent (iii) The law of the random variable $D_{\kappa}$ manifests a large deviation property with 
parameter $\kappa^{d-2}$.

\smallskip

\noindent For the normal phase:

\noindent (i) also holds; (ii) holds but for $\kappa^{d/2}(D_{\kappa} -m)$, instead of
$\kappa^{(d-2)/2}(D_{\kappa} -m)$; and (iii) holds with the order
$\kappa^{d}$, instead of $\kappa^{d-2}$.

The comparison shows that there are differences in deviation of density fluctuation
between the BEC and the non-BEC states of ideal boson gases, which reminds the
large deviation properties for two-phase classical systems, for example lattice spin models, see e.g. \cite{P}.
The specificity of the BEC is that it is a \textit{quantum phase transition} with particular quantum
fluctuations \cite{LePu}, \cite{ZB}.

%%%%%%%%%%%%%%%%%%%%%%%%%%%%%%%%% Acknowledgements %%%%%%%%%%%%%%%%%%%%%%%%%%%%%%%%%%%%%%%%
\bigskip

\noindent \textbf{Acknowledgments}

\noindent H.T. thanks JSPS for the financial support
under the Grant-in-Aid for Scientific Research (C) 20540162
and Centre de Physique Th\'{e}orique, Luminy-Marseille for hospitality.
V.A.Z. is grateful to Mathematical Department of the Kanazawa University for a warm
hospitality and for financial support.

%%%%%%%%%%%%%%%%%%%%%%%%%%%%%%%%%%%%%%%%%%%%%%%%%%%%%%%%%%%%%%%%%%%%%%%

\end{document}